# Implementation of annihilation and junction reactions in vector density-based continuum dislocation dynamics

**Peng Lin and Anter El-Azab**

School of Materials Engineering, Purdue University, West Lafayette, IN 47907, USA

Email: lin936@purdue.edu, aelazab@purdue.edu

**Abstract**

In a continuum dislocation dynamics formulation by Xia and El-Azab [1], dislocations are represented by a set of vector density fields, one per crystallographic slip systems. The space-time evolution of these densities is obtained by solving a set of dislocation transport equations coupled with crystal mechanics. Here, we present an approach for incorporating dislocation annihilation and junction reactions into the dislocation transport equations. These reactions consume dislocations and result in nothing as in the annihilation reactions, or produce new dislocations of different types as in the case of junction reactions. Collinear annihilation, glissile junctions, and sessile junctions are particularly emphasized here. A generalized energy-based criterion for junction reactions is established in terms of the dislocation density and Burgers vectors of the reacting species, and the reaction rate terms for junction reactions are formulated in terms of the dislocation densities. In order to illustrate how the dislocation network changes as a result of junction formation and annihilation in a continuum dislocation dynamics setting, we present some numerical examples focusing on the reactions processes themselves. The results show that our modeling approach is able to capture the respective dislocation network changes associated with dislocation reactions in FCC crystals: dislocations of opposite line directions encountering each other on collinear slip systems annihilate to connect the dislocations on the two slip systems, glissile junctions form on new slip system behave like Frank-Read sources, and sessile junctions form and expand along the intersection of the slip planes of the reacting dislocation species. A collective-dynamics test showing the frequency of occurrence of junctions of different types relative to each other is also presented.

## 1. Introduction

The plastic strength of metallic crystals derives from the motion, multiplication and reactions of dislocations at the mesoscale. Since dislocations were theoretically postulated [2] and confirmed by experiment [3], much of the research on dislocation in structural metals has focused on dislocation interactions and their role in strain hardening. When dislocations on different slip systems interact with each other, different kinds of junctions are formed, a process that depends on the Burgers vectors and slip planes of the reacting dislocations. Discrete Dislocation Dynamics (DDD) simulations [4] showed that the dislocation network formed by different kinds of junctions is the microstructural origin of strain hardening and that different junctions have different contributions to the overall hardening. Junction formation also contributes to the multiplication of dislocations by pinning or constraining the end points of junctions causing the rest of the dislocation lines to bow out. For example, glissile junctions act as Frank-Read (FR)-like source with endpoints free to move along the intersection line of glide planes, contributing significantly to overall density changes [5,6]. In DDD, junctions are implemented as topological rearrangement of the dislocation network given certain criteria of formation [6–8]. In continuum crystal plasticity, on the other hand, Taylor hardening terms are incorporated into the formulation to consider the influence of dislocation junctions. The resulting hardening terms are often considered to be proportional to the square root of the forest dislocation density [9–12]. As proposed in [13], such terms are expressed in terms of the sum of the dislocation densities on forest slip systems weighted by the strength of the corresponding interactions. The interaction coefficients have different values for different types of junctions, representing the average strength of the mutual interactions between the two interacting slip systems. These interaction coefficients values have been calculated by DDD simulations [14–17]. Another approach that includes the glissile junction formation into the rate equations for the dislocation density evolution has also been adopted in a recent crystal plasticity model [18].

Aiming to capture the collective behavior of dislocations, Continuum Dislocation Dynamics (CDD) must incorporate some representation of dislocation annihilation and junction reactions. Recent coarse graining approaches based on statistical mechanics [19–21] have shown that Taylor hardening-like terms naturally appear in parallel dislocation systems of the same Burgers vector when the dislocation-dislocation correlations are considered. As such, at the mesoscale, the scale immediately above the discrete dislocation scale, Taylor type hardening is not necessarily associated with junctions and some sort of such terms should exist in addition to the explicit representation of junctions. At the macroscale, however, Taylor type terms are believed to be suitable for describing the overall response of crystals. On the other hand, the change of dislocation network due to junction reaction is not considered by Taylor type terms, which can play an important role in forming dislocation microstructure at the mesoscale. For example, the formation of a glissile junction amounts to the reaction of two dislocations to generate a dislocation segment on a third slip system. This process thus involves dislocation exchange among different slip systems. Also, when one dislocation encounters another, they can either form a junction or a jog associated with forest cutting, with the outcome depending on the angle between the two dislocations. So, the line directions of the reacting dislocations should also be considered in sorting out these reactions. Having said so, the method used to account for the orientations of the reacting dislocations in DDD is not directly applicable to the case of continuum dislocation dynamics because, in the latter, dislocations are represented by continuum density-like variables instead of discrete line segments. A reformulation of the topological treatment of dislocation network should thus be considered in the case of continuum representation of dislocations. Inspired by their DDD results showing glissile junctions acting as dislocation sources, Stricker and Weygand [5] suggested a rate term of glissile junction formation to be included in the CDD framework. Striker et al [22] further emphasized the importance of including dislocation density coupling between slip systems due to dislocation network rearrangement, which was motivated by finding multiplication to be a more dominant dislocation density evolution than transport in stage II deformation.

In the last two decades, several attempts have been made to formulate density-based models for the evolution of dislocation microstructures based on statistical mechanical concepts. Pioneering models were established for systems of parallel straight dislocations in two dimensions (2D) by Groma and Balogh [23], Zaiser et al [19], Groma et al [20], Rodney et al [24] and Kooiman et al [25]. The coupled evolution of total dislocation density and the net signed dislocation density, also known as the geometric density, can be captured by these models. However, in these 2D CDD models, it is quite difficult to consider the junction formation explicitly. Extending the 2D approach to 3D, where the dislocations are modeled as curved lines moving perpendicular to their line direction in their slip planes, has proven to be quite challenging. Different approaches have been made to represent 3D dislocation configurations. A 7D phase space $\mathbb{R}^3 \times \mathbb{R}^3 \times \mathbb{T}$ was used to characterize 3D curved dislocations, where $\mathbb{R}^3$ is the 3D Euclidean space and $\mathbb{T}$ is the orientation defining the local line tangent of the dislocation in their slip planes [26–28]. A scalar density $\rho(\mathbf{x},t)$ with unit line direction $\xi(\mathbf{x},t)$ was also used in other model [29]. Some other authors expressed the dislocation configuration into screw density $\rho_{\text{screw}}$ and edge density $\rho_{\text{edge}}$ [30–32]. Another approach [33] treats the 3D curved dislocation lines in a higher dimensional space containing line orientation variables as extra dimensions, so densities can carry additional information about their line direction and curvature. Simplified variants of the latter theory have been formulated, which consider only low-order moments of the dislocation orientation distribution [34,35]. One further development of this theory is achieved through a hierarchy of evolution equations of the so-called alignment tensor, which contains information on the directional distribution of dislocation density and dislocation curvature. Although these models have successfully described dislocation transport, dislocation reactions between different slip systems were incorporated into them only recently, for example, by associating dislocation multiplication and annihilation with changes in the volume density of dislocation loops—see the recent works by Monavari and Zaiser [36] and Sudmanns et al [37] incorporating dislocation junctions and annihilation into a CDD framework.

In this paper, the vector-density based formulation of Xia et al [1,38] is considered as a starting point. In this formulation, the dislocations are represent by a set of vector fields, $\boldsymbol{\rho}^{(\alpha)}$, where $\alpha = 1,\cdots,N$, with $N$ being the number of slip systems. In this representation, the dislocations on a given slip system are considered to be bundles with a unique line direction at each point in space, and the magnitude of the corresponding density field represents the local scalar density of dislocations in the bundle. Kinetic equations were established to describe the space and time evolution of the dislocation densities on each slip system, with transport and reactions being the main evolution mechanisms. The reactions include the cross slip as a simple transfer of dislocation among collinear systems sharing the same Burgers vector, annihilation of dislocations of the same Burgers vector and opposite line directions, and the formation of junctions. The latter reactions differ in that they consume two species of dislocations and produce a third species. If the product species is mobile (glissile), it is assigned to the third slip system. Immobile (sessile) junctions on the other hand are considered but not assigned to regular slip systems. All reactions are described here by coupling terms in the kinetic equations governing the evolution of dislocations.

In section 2, the vector-density based CDD model is briefly introduced together with the mechanisms which cause dislocation network change. In section 3, the criterion for collinear annihilation and the energy-based criteria for junction reactions are explained. In section 4, the coupling terms for dislocation evolution due to junction formation are derived. In section 5, several test problems are presented, followed by a discussion section (section 6) regarding different rate formulas. Closing remarks are made in section 7.

## 2. Continuum dislocation dynamics with vector dislocation densities

We begin by a brief introduction of the CDD model under consideration. In this model, the dislocations on a given slip system are represented by a vector density field $\boldsymbol{\rho}^{(\alpha)} = \rho^{(\alpha)} \boldsymbol{\xi}^{(\alpha)}$, where $\boldsymbol{\xi}^{(\alpha)}$ is the line direction of the dislocation bundle and $\rho^{(\alpha)}$ is the scalar density of dislocations. The evolution of density field $\boldsymbol{\rho}^{(\alpha)}$ is described by a transport equation of the form[1,38]

$$\dot{\boldsymbol{\rho}}^{(\alpha)} = \nabla \times \left( \mathbf{v}^{(\alpha)} \times \boldsymbol{\rho}^{(\alpha)} \right), \quad \alpha = 1, \cdots, N, \tag{1}$$

where $\mathbf{v}^{(\alpha)}$ is the velocity of the dislocation bundle. Equation (1) is valid for dislocations on the same slip system. For the multiple slip case, a system of transport equations of the form (1) are to be solved concurrently for the space and time evolution of dislocations on all slip systems.

The solution of the system (1) requires the velocity field $\mathbf{v}^{(\alpha)}$ as input. In the CDD model under consideration, the dislocation velocity is fixed by evaluating the internal stress field from which the Peach-Koehler force on each slip system is evaluated and then used to fix the corresponding velocity via a dislocation mobility law. The internal stress of the dislocations is calculated by solving the eigenstrain boundary value problem:

$$\begin{cases} \nabla \cdot \boldsymbol{\sigma} = \mathbf{0} & \text{in } \Omega \\ \boldsymbol{\sigma} = \mathbf{C} : (\nabla \mathbf{u} - \boldsymbol{\beta}^{\mathrm{p}}) & \text{in } \Omega \\ \mathbf{u} = \bar{\mathbf{u}} & \text{on } \partial \Omega_u \\ \mathbf{n} \cdot \boldsymbol{\sigma} = \bar{\mathbf{t}} & \text{on } \partial \Omega_\sigma \end{cases}, \tag{2}$$

where $\boldsymbol{\sigma}$ is the Cauchy stress, $\mathbf{C}$ is the symmetric, forth rank elastic tensor, $\mathbf{u}$ is the displacement field, $\boldsymbol{\beta}^{\mathrm{p}}$ is the plastic distortion tensor, $\mathbf{n}$ is the unit normal to the boundary $\partial \Omega$, and $\bar{\mathbf{u}}$ and $\bar{\mathbf{t}}$ are the displacement and traction boundary conditions, respectively. Formally speaking, Cauchy stress in equation (2) is equal to the elastic tensor times the symmetric part of the elastic distortion, $\boldsymbol{\beta}^e = (\nabla \mathbf{u} - \boldsymbol{\beta}^p)$. However, the form of Hooke's law (2) is still valid since the elastic tensor is symmetric. The plastic distortion is determined by summing the plastic slip over all slip systems,

$$\boldsymbol{\beta}^p = \sum_\alpha \gamma^{(\alpha)} \mathbf{m}^{(\alpha)} \otimes \mathbf{s}^{(\alpha)}, \tag{3}$$

where $\mathbf{m}^{(\alpha)}$ is the unit normal vector of the slip plane of slip system $\alpha$, $\mathbf{s}^{(\alpha)}$ is its unit slip vector (along Burgers vector), and $\gamma^{(\alpha)}$ is the corresponding crystal slip. The dislocation glide velocity on a given slip system is assumed to change linearly with the local resolved shear stress on that slip system [19,39],

$$v^{(\alpha)} = \frac{b^{(\alpha)}}{B} \left\langle \left|\tau^{(\alpha)}\right| - \left(\tau_0 + \tau_p\right) \right\rangle \operatorname{sgn}\left(\tau^{(\alpha)}\right), \tag{4}$$

where $b^{(\alpha)}$ is the magnitude of Burgers vector and $B$ is the drag coefficient. The sign function returns the signature of its argument and $\langle \cdot \rangle$ denote the Macaulay brackets, which return the argument if it is positive and zero otherwise. $\tau_0$ is the stress representing lattice friction and $\tau^{(\alpha)}$ is the resolved shear stress, $\tau^{(\alpha)} = \mathbf{m}^{(\alpha)} \cdot \boldsymbol{\sigma} \cdot \mathbf{s}^{(\alpha)}$. Cauchy stress $\boldsymbol{\sigma}$ accounts for the combination of long-range interaction stress of dislocations and the stress arising due the imposed boundary conditions. The Taylor hardening stress $\tau_p$ accounts for the short-range interactions due to sessile dislocation junction reactions and jog formation by the cutting of forest dislocations. Typically, this term has the following form [13,16,40]:

$$\tau_p^{(\alpha)} = \mu b \sqrt{\sum_\beta a^{\alpha\beta} \rho^{(\beta)}}, \tag{5}$$

with $\mu$ being the shear modulus and $a^{\alpha\beta}$ the interaction matrix. Statistical modeling has shown that Taylor-like friction stress terms arise from the dislocation-dislocation correlation, which are found to be short ranged in the idealized long, parallel straight dislocations [19,20]. A thermodynamic treatment within the CDD framework reported in [41] also shows that such terms are possible in CDD. In the expression (5), the density of dislocations interacting at short range with dislocations on a given slip system $\alpha$ is split into the sum of the reacting densities weighted by the strength of the corresponding interactions. The corresponding coefficients $a^{\alpha\beta}$ represent the average strength of the mutual interactions between slip systems $\alpha$ and $\beta$. For symmetry considerations, the number of distinct interaction coefficients between 12 mutually interacting slip systems in FCC crystal is reduced to only six, which are associated with the self, coplanar and collinear interactions, and the formation of glissile junctions, Lomer locks, and Hirth locks [14–17].

By adding the Taylor hardening term, dislocations will slow down where there are junctions. However, the change of dislocation network, i.e., change of the connectivity of dislocations on different slip system, cannot be captured. For example, a glissile junction formed by two dislocation segments can glide within a third slip system. Hence, in order to have a more accurate description of the dislocation density evolution, explicit coupling terms associated with junction reactions should be added to equation (1). Not only so, but such terms should also account for cross slip and annihilation. We thus rewrite the transport equation (1) in the form:

$$\dot{\boldsymbol{\rho}}^{(\alpha)} = \nabla \times (\mathbf{v}^{(\alpha)} \times \boldsymbol{\rho}^{(\alpha)}) + \dot{\boldsymbol{\rho}}_{cp}^{(\alpha)}, \quad \alpha = 1, \cdots, N, \tag{6}$$

where $\dot{\boldsymbol{\rho}}_{cp}^{(\alpha)}$ is a coupling term, which is the time rate of change of $\boldsymbol{\rho}^{(\alpha)}$ due to cross slip, collinear annihilation, and junction reactions. These mechanisms, which are illustrated in figure 1, will cause the dislocation network to change. (a) *Cross slip*: A screw dislocation can move from one slip plane to another to avoid a barrier on its initial slip plane. (b) *Collinear annihilation*: Two anti-parallel screw dislocations initially gliding on different slip planes and having the same Burgers vector will annihilate when they encounter each other at the intersection of their slip planes. (c) *Glissile junction*: The formation of a dislocation junction involves two dislocation segments on two different slip systems. In the case of a glissile junction, let us assume that the Burgers vector of the first slip system $\mathbf{b}^{(1)}$ is parallel to the intersection line of the two slip planes of reacting dislocations. For a combination of directions of the two dislocation lines which leads to an attractive elastic interaction, the junction formed is glissile on the second slip plane for purely geometrical reasons and characterized by $(\mathbf{b}^{(1)} + \mathbf{b}^{(2)}, \mathbf{m}^{(2)})$. The result of the junction is a mobile dislocation on the new slip system, where $\mathbf{b}^{(\alpha)}$ and $\mathbf{m}^{(\alpha)}$, respectively, refer to the Burgers vector and slip plane normal of the dislocations involved. (d) *Sessile junction*: Similar to the case of a glissile junction, dislocations on the reacting slip systems will be consumed in the process with the resulting junction stuck in place along the intersection of the two slip systems of the reacting dislocations.

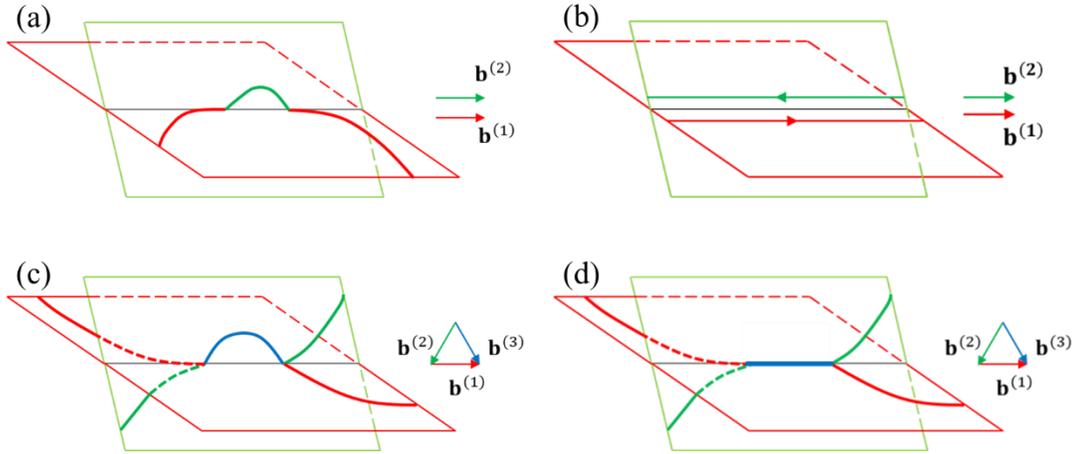

Figure 1. Dislocation network changes due to dislocation reactions: (a) cross slip, (b) collinear annihilation, (c) glissile junction, and (d) sessile junction.

The coupling term $\dot{\rho}_{cp}^{(\alpha)}$ in equation (6) can be established once the rates of the above reactions are formulated. In order to do so, criteria for the dislocation reactions should first be established so as to know when and where to activate such reactions. Also, the corresponding coupling terms should be established in terms of vector densities of dislocations involved so that the system of equations (6) becomes self-consistent. In the following two sections, we will discuss these topics in detail. Cross slip in continuum dislocation dynamics has been modelled in some earlier works [1,36,38]. Hence, we here present models for only collinear annihilation, glissile junction and sessile junction in continuum dislocation dynamics.

## 3. Junction reactions and related criteria

### *3.1. Burgers vector considerations*

When two dislocations form a junction, the resulting dislocation lies on the intersection of the slip planes of the reacting dislocations, with the junction Burgers vector being the sum of the two Burgers vectors of the reacting ones. The plane defined by the junction dislocation line and its Burgers vector determines whether it is glissile or sessile. If the plane coincides with a slip plane of the crystal, the resulting junction is glissile, and it is sessile otherwise. The collinear annihilation is considered here as a type of junction with a zero Burgers vector of the product segment. Hence, we have three types of junctions, for which Burgers vectors should satisfy the conditions,

$$\begin{aligned}
\mathbf{b}^{(1)} &= \mathbf{b}^{(2)}, &&\text{collinear annihilation} \\
\mathbf{b}^{(1)} + \mathbf{b}^{(2)} &= \mathbf{b}^{(3)}, &&\text{glissile junction} \\
\mathbf{b}^{(1)} + \mathbf{b}^{(2)} &= \mathbf{b}_{LC} \text{ or } \mathbf{b}_{H}, &&\text{sessile junction}
\end{aligned} \qquad (7)$$

Here, $\mathbf{b}^{(\alpha)}$ are Burgers vectors of the primary slip systems, and $\mathbf{b}_{LC}$ and $\mathbf{b}_{H}$ are Burgers vectors of the Lomer-Cottrell and Hirth junctions, respectively. For a FCC crystal, the 12 slip systems are defined as table 1 and possible types of junctions are listed in table 2 and table 3. Lomer-Cottrell junctions have Burgers vector of <110> type and the glide plane of {100} type, and are thus sessile. Hirth junctions have Burgers vector of <200> type, which is not a slip vector. It should be pointed out that (see table 2) junctions arising from different reactions may have the same line direction and Burgers vector. For example, junction 1 and junction 2 have the same line direction and Burgers vector; however, the former is formed among slip systems 2 and 7, while the latter is formed among slip systems 3 and 6. By considering Burgers vectors only, all types of junction reactions are listed in table 4. For a given slip system, there can be one collinear annihilation, four glissile junctions, two Lomer-Cottrell junctions, and two Hirth junctions.

Table 1. Primary slip systems of FCC crystal.

| No. | 1 | 2 | 3 | 4 | 5 | 6 | 7 | 8 | 9 | 10 | 11 | 12 |
|---|---|---|---|---|---|---|---|---|---|---|---|---|
| Slip plane | (111) | ($\bar{1}$11) | ($\bar{1}$11) | ($\bar{1}\bar{1}$1) | ($\bar{1}\bar{1}$1) | (1$\bar{1}$1) | (1$\bar{1}$1) | (111) | (111) | ($\bar{1}\bar{1}$1) | ($\bar{1}$11) | (1$\bar{1}$1) |
| Slip direction | [0$\bar{1}$1] | [0$\bar{1}$1] | [101] | [101] | [011] | [011] | [$\bar{1}$01] | [$\bar{1}$01] | [$\bar{1}$10] | [$\bar{1}$10] | [$\bar{1}\bar{1}$0] | [$\bar{1}\bar{1}$0] |

Table 2. The line direction $\tilde{\mathbf{e}}$ and type of Burgers vector $\mathbf{b}_{LC}$ of Lomer-Cottrell junctions.

| No. | 1 | 2 | 3 | 4 | 5 | 6 | 7 | 8 | 9 | 10 | 11 | 12 |
|---|---|---|---|---|---|---|---|---|---|---|---|---|
| $\sqrt{2}\,\tilde{\mathbf{e}}$ | [110] | [110] | [101] | [101] | [011] | [011] | [1$\bar{1}$0] | [1$\bar{1}$0] | [10$\bar{1}$] | [10$\bar{1}$] | [01$\bar{1}$] | [01$\bar{1}$] |
| $\mathbf{b}_{LC}$ type | [1$\bar{1}$0] | [1$\bar{1}$0] | [10$\bar{1}$] | [10$\bar{1}$] | [01$\bar{1}$] | [01$\bar{1}$] | [110] | [110] | [101] | [101] | [011] | [011] |

Table 3. The line direction $\tilde{\mathbf{e}}$ and type of Burgers vector $\mathbf{b}_H$ of Hirth junctions.

| No. | 1 | 2 | 3 | 4 | 5 | 6 | 7 | 8 | 9 | 10 | 11 | 12 |
|---|---|---|---|---|---|---|---|---|---|---|---|---|
| $\sqrt{2}\,\tilde{\mathbf{e}}$ | [110] | [110] | [101] | [101] | [011] | [011] | [1$\bar{1}$0] | [1$\bar{1}$0] | [10$\bar{1}$] | [10$\bar{1}$] | [01$\bar{1}$] | [01$\bar{1}$] |
| $\mathbf{b}_H$ type | [020] | [200] | [002] | [200] | [002] | [020] | [020] | [200] | [002] | [200] | [002] | [020] |

Table 4. Types of junctions formed by different slip systems. "col" refers to as collinear annihilation; "g" refers to as glissile junction; "LC" refers to as Lomer-Cottrell junction; "H" refers to as Hirth junction. The number in parentheses is the glissile junction number in table 1 or sessile junction defined in table 2 and table 3. It is to be noted that glissile junctions belong to the set of primary slip systems shown in table 1.

|    | 1 | 2 | 3 | 4 | 5 | 6 | 7 | 8 | 9 | 10 | 11 | 12 |
|----|---|---|---|---|---|---|---|---|---|----|----|----|
| 1  |   | col | g(11) | LC(7) | H(7) | H(9) | g(9) |   |   | g(8) | g(3) | LC(9) |
| 2  | col |   |   | g(11) | H(3) | H(1) | LC(1) | g(9) | g(8) | LC(3) |   | g(3) |
| 3  | g(11) |   |   | col | g(10) | LC(2) | H(2) | H(11) | LC(11) | g(5) |   | g(2) |
| 4  | LC(7) | g(11) | col |   |   | g(10) | H(5) | H(8) | g(5) |   | g(2) | LC(5) |
| 5  | H(7) | H(3) | g(10) |   |   | col | g(12) | LC(8) | g(4) |   | LC(4) | g(7) |
| 6  | H(9) | H(1) | LC(2) | g(10) | col |   |   | g(12) | LC(10) | g(4) | g(7) |   |
| 7  | g(9) | LC(1) | H(2) | H(5) | g(12) |   |   | col | g(1) | LC(6) | g(6) |   |
| 8  |   | g(9) | H(11) | H(8) | LC(8) | g(12) | col |   |   | g(1) | LC(12) | g(6) |
| 9  |   | g(8) | LC(11) | g(5) | g(4) | LC(10) | g(1) |   |   | col | H(12) | H(10) |
| 10 | g(8) | LC(3) | g(5) |   |   | g(4) | LC(6) | g(1) | col |   | H(4) | H(6) |
| 11 | g(3) |   | g(2) | LC(4) | g(7) | g(6) | LC(12) | H(12) | H(4) |   | col |   |
| 12 | LC(9) | g(3) | g(2) | LC(5) | g(7) |   | g(6) | H(10) | H(6) | col |   |   |

### *3.2. Line direction considerations*

Table 4 show the possible junction reactions among different slip systems by Burgers vector considerations alone. However, whether a given junction reaction actually happens depends on the line directions of dislocations. Dislocations crossing each other at, say, vertical angle do not form junctions. Instead, they are more likely to pass each other, leaving jogs on the dislocation lines. Dislocation line direction considerations are thus important.

For a collinear annihilation to happen, the two dislocations should be in opposite directions. In the present work, the criterion for collinear annihilation reaction is expressed in the form

$$\pi - \varphi_c \leq \varphi \leq \pi, \tag{8}$$

where $\varphi = \arccos(\boldsymbol{\rho}^{(1)} \cdot \boldsymbol{\rho}^{(2)} / \rho^{(1)} \rho^{(2)})$ is the angle between the two dislocation lines, as shown in figure 2 and $\varphi_c$ is a material parameter chosen to be $\pi/12$ in our numerical implementation. The choice of this parameter is based on our previous CDD simulations for cross slip, which was adopted based on discrete dislocation dynamics considerations [1,38]. Here we assume the collinear annihilation share a similar orientation dependence.

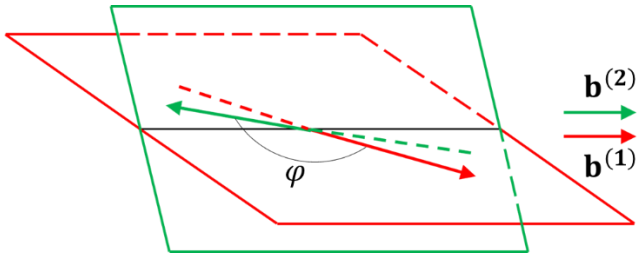

Figure 2. A schematic showing two dislcoation lines prior to a collinear annhilation reaction. The Burgers vectors are of the same type and the line directions are nearly opposite.

For junction reactions between dislocations on two different slip planes, the angles between the parent segments and the intersection of the two slip planes determine whether the junction can form. This line orientation dependence is valid for both sessile and glissile junctions. An energy-based criterion can be established to study this situation [42–44]. It is commonly accepted that the energy associated with a dislocation is mainly contributed by the elastic energy associated with the long-range elastic strain of the dislocation, while other contributions such as core energy are generally neglected [45]. The classical expression of elastic energy $E$ per unit length of a straight dislocation with mixed character in an isotropic linear elastic crystal is given by [46]

$$E(\beta) = \frac{\mu b^2 \left(1 - \nu \cos^2 \beta\right)}{4\pi(1-\nu)} \ln\left(\frac{R}{r_0}\right), \tag{9}$$

where $\beta$ is the angle between the Burgers vector $\mathbf{b}$ and the dislocation line tangent vector $\xi$, $R$ and $r_0$ are, respectively, the outer and inner cut-off radii, $b = |\mathbf{b}|$, and $\mu$ and $v$ are the shear modulus and Poisson ratio. The energy of a dislocation with a short length $l$ is written in the form

$$E(\beta,l) = \varsigma\mu b^2 \left(1 - v\cos^2\beta\right)l, \tag{10}$$

with $\varsigma = \ln(R/r_0)/4\pi(1-v)$ being a material constant. After forming a small junction segment $dl_j$, the lengths of the two parent segments decrease and their line direction also changes, as shown in figure 3. The variation of total dislocation line energy can be written as

$$dE = \varsigma\mu b^2 \left[\left(1 - v\cos^2\beta\right)dl + v\sin(2\beta)l\,d\beta\right]. \tag{11}$$

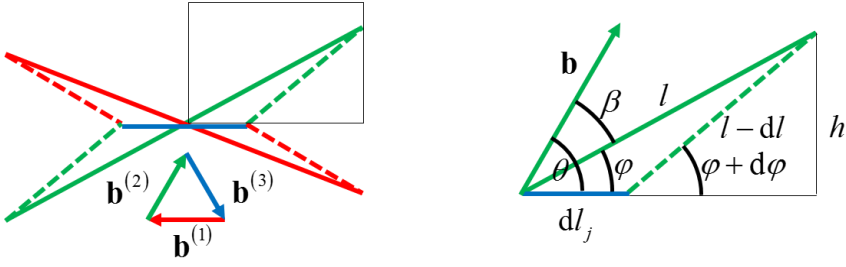

Figure 3. A geometric illustration of a junction reaction according to [44]. The red and the green solid lines are the dislcoation lines prior to junction formation. The dashed lines represent the configuration after forming the short blue junction segment. The part of the confiurtion within the box is shown to the right of the figure in detail.

From figure 3, we have the following geometric relation:

$$l\sin\varphi = h, \quad \beta = \theta - \varphi. \tag{12}$$

The differential changes $dl$, $d\varphi$ and $d\beta$ in $l$, $\varphi$ and $\beta$ can be written in terms of junction segment length $dl_j$ as below:

$$dl = -\cos\varphi\, dl_j, \quad d\varphi = -\frac{dl\tan\varphi}{l}, \quad d\beta = -d\varphi. \tag{13}$$

By substituting Eq. (13) into Eq. (11) and summing over all three slip systems, the change of energy by forming a differential junction segment can be derived as

$$dE_j = \varsigma\mu b^2 \left\{\left(1 - v\cos^2\beta^{(3)}\right) - \sum_{k=1,2}\left[\left(1 - v\cos^2\beta^{(k)}\right)\cos\varphi^{(k)} + v\sin(2\beta^{(k)})\sin\varphi^{(k)}\right]\right\}dl_j. \tag{14}$$

The criterion for forming the junction is then: $dE_j < 0$. Unlike Frank criterion for dislocation reactions [46,47], which accounts only for Burgers vector, this energy criterion contains all details of the dislocation configuration. This energy expression (14) was previously derived by Madec eta al [44,48] in the context of junction implementation in discrete dislocation dynamics. It is included here for completeness and to facilitate the generalization made in the current work and the extension to continuum dislocation dynamics.

A generalized form of the energy change with junction formation suitable for implementation with all kinds of junctions can be established by a vector representation of the junction configuration. Let us first rewrite the energy per unit length of a discrete dislocation line equation (9), in terms of its Burgers vector and its line tangent. For a dislocation of Burgers vector $\mathbf{b}$ and line tangent $\boldsymbol{\xi}$, the energy of dislocation per unit length can be written, as a generalization of equation (9), in the form

$$E(\mathbf{b},\boldsymbol{\xi}) = \varsigma\mu |\mathbf{b}|^2 \left(1 - \nu \left(\frac{\mathbf{b}\cdot\boldsymbol{\xi}}{|\mathbf{b}||\boldsymbol{\xi}|}\right)^2\right) |\boldsymbol{\xi}|, \tag{15}$$

where $|\mathbf{x}|$ is the norm of the vector $\mathbf{x}$, and $|\boldsymbol{\xi}|$, which is unity, is kept in the equation as a place holder. This expression can be generalized to a density based representation of dislocation by replacing the line tangent $\boldsymbol{\xi}$ with the vector density $\boldsymbol{\rho}^{(\alpha)}$

$$E^{(\alpha)}\left(\mathbf{b}^{(\alpha)},\boldsymbol{\rho}^{(\alpha)}\right) = \varsigma\mu |\mathbf{b}^{(\alpha)}|^2 \left(1 - \nu \left(\frac{\mathbf{b}^{(\alpha)}\cdot\boldsymbol{\rho}^{(\alpha)}}{|\mathbf{b}^{(\alpha)}||\boldsymbol{\rho}^{(\alpha)}|}\right)^2\right) |\boldsymbol{\rho}^{(\alpha)}|. \tag{16}$$

This expression represents energy density, i.e., energy per unit volume since $\boldsymbol{\rho}^{(\alpha)} = \rho^{(\alpha)}\boldsymbol{\xi}^{(\alpha)}$ and $\rho^{(\alpha)}$ is the scalar dislocation density. An energy criterion for junction formation in continuum dislocation dynamics can now be derived starting from the last expression. Assuming that a differential continuum junction density $\Delta\boldsymbol{\rho}_{\text{junc}}^{(1,2)} = \Delta\rho_{\text{junc}}^{(1,2)}\tilde{\mathbf{e}}$ is formed at a point in space, where $\tilde{\mathbf{e}}$ is a unit vector indicating junction direction, then according to equation (16) the total energy after the junction configuration can be written as

$$\begin{aligned}
E_{\text{total}}&(\mathbf{b}^{(1)},\boldsymbol{\rho}^{(1)},\mathbf{b}^{(2)},\boldsymbol{\rho}^{(2)},\Delta\rho_{\text{junc}}^{(1,2)}\tilde{\mathbf{e}}) \\
&= E^{(1)}(\mathbf{b}^{(1)},\boldsymbol{\rho}^{(1)} - \Delta\rho_{\text{junc}}^{(1,2)}\tilde{\mathbf{e}}) + E^{(2)}(\mathbf{b}^{(2)},\boldsymbol{\rho}^{(2)} - \Delta\rho_{\text{junc}}^{(1,2)}\tilde{\mathbf{e}}) + E^{(3)}(\mathbf{b}^{(1)} + \mathbf{b}^{(2)},\Delta\rho_{\text{junc}}^{(1,2)}\tilde{\mathbf{e}})
\end{aligned} \tag{17}$$

Here, we made use of the fact that $\mathbf{b}^{(3)} = \mathbf{b}^{(1)} + \mathbf{b}^{(2)}$. In the last expression, $\Delta \rho_{\text{junc}}^{(1,2)} = 0$ prior to the formation of the junction. The condition for establishing the junction is that the total energy decreases upon its formation, which means the following,

$$\left. \frac{\partial E_{\text{total}}}{\partial \Delta \rho_{\text{junc}}^{(1,2)}} \right|_{\Delta \rho_{\text{junc}}^{(1,2)} = 0} < 0 \; . \tag{18}$$

By using equation (16), we then obtain

$$\left. \frac{\partial E^{(k)}(\mathbf{b}^{(k)}, \boldsymbol{\rho}^{(k)} - \Delta \rho_{\text{junc}}^{(1,2)} \tilde{\mathbf{e}})}{\partial \Delta \rho_{\text{junc}}^{(1,2)}} \right|_{\Delta \rho_{\text{junc}}^{(1,2)} = 0}$$
$$= -\varsigma \mu (\boldsymbol{\xi}^{(k)} \cdot \tilde{\mathbf{e}})(|\mathbf{b}^{(k)}|^2 - \nu (\mathbf{b}^{(k)} \cdot \boldsymbol{\xi}^{(k)})^2) + 2 \varsigma \mu \nu ((\mathbf{b}^{(k)} \cdot \tilde{\mathbf{e}})(\mathbf{b}^{(k)} \cdot \boldsymbol{\xi}^{(k)}) - (\boldsymbol{\xi}^{(k)} \cdot \tilde{\mathbf{e}})(\mathbf{b}^{(k)} \cdot \boldsymbol{\xi}^{(k)})^2) \tag{19}$$

for $k = 1, 2$, with $\boldsymbol{\xi}^{(k)} = \boldsymbol{\rho}^{(k)} / \rho^{(k)}$, and

$$\left. \frac{\partial E^{(3)}(\mathbf{b}^{(1)} + \mathbf{b}^{(1)}, \Delta \rho_{\text{junc}}^{(1,2)} \tilde{\mathbf{e}})}{\partial \Delta \rho_{\text{junc}}^{(1,2)}} \right|_{\Delta \rho_{\text{junc}}^{(1,2)} = 0} = \varsigma \mu (|\mathbf{b}^{(1)} + \mathbf{b}^{(2)}|^2 - \nu ([\mathbf{b}^{(1)} + \mathbf{b}^{(2)}] \cdot \tilde{\mathbf{e}})^2) \; . \tag{20}$$

So the energy criterion can be written as a function of reacting dislocation density vectors, $\boldsymbol{\rho}^{(1)}$ and $\boldsymbol{\rho}^{(2)}$, the corresponding Burgers vectors, $\mathbf{b}^{(1)}$ and , together with the line direction of the junction $\tilde{\mathbf{e}}$, which is a constant vector for each type of junction, by substituting equations (19) and (20) into equation (18). When using the criterion (18), the dislocation density vectors $\boldsymbol{\rho}^{(k)}$ should be consistent with the direction of the intersection vector, which means $\boldsymbol{\rho}^{(k)} \cdot \tilde{\mathbf{e}} > 0$. If not, we can change the sign of both the density $\boldsymbol{\rho}^{(k)}$ and its Burgers vector $\mathbf{b}^{(k)}$ to make it satisfied, since the physical dislocation does not change by changing the sign of $\boldsymbol{\rho}^{(k)}$ and $\mathbf{b}^{(k)}$ simultaneously . Figure 4 shows the range of $\varphi^{(1)}$ and $\varphi^{(2)}$ where junction reaction can happen, where $\varphi^{(1)}$ (or $\varphi^{(2)}$) is the angle between dislocation $\boldsymbol{\rho}^{(1)}$ (or $\boldsymbol{\rho}^{(2)}$) and the intersection. Taking glissile junction as an example, it can be seen that the region is symmetric about $\varphi^{(1)}$, but asymmetric about $\varphi^{(2)}$, because $\mathbf{b}^{(1)}$ is parallel to the intersection of the two slip systems while $\mathbf{b}^{(2)}$ is not.

We remark that the energy criterion derived here applies to all kinds of junctions. In the case of junctions forming between dislocations initially on two different slip systems, the unit junction direction $\tilde{\mathbf{e}}$ falls along the line of intersection of the two slip planes. In the case of a glissile junction forming between dislocations on the same slip plane, that line direction can be simply taken to be the middle direction between the reacting directions (ignoring the effects of local curvature).

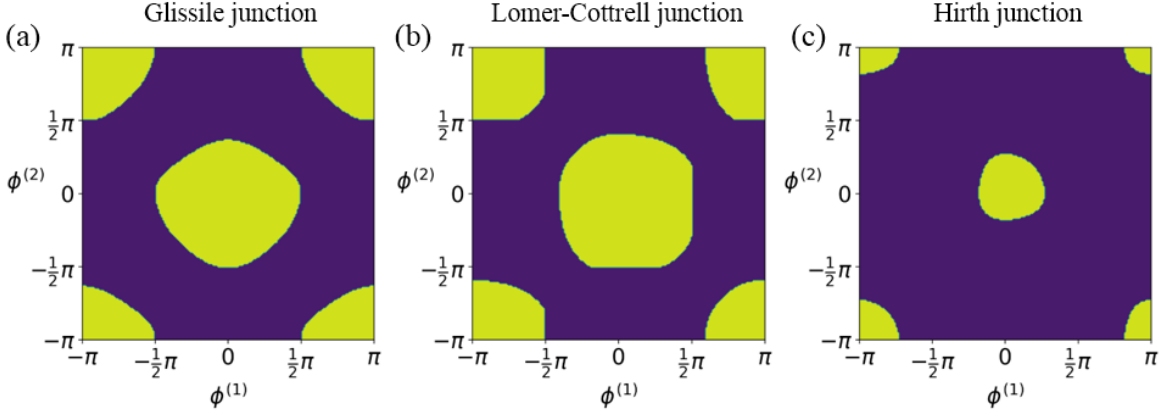

Figure 4. Orientation dependence of different junction reactions calculated by the energy criterion (equation (18)). The two interacting slip systems are: (a) $(111)[01\bar{1}]$ and $(\bar{1}11)[101]$ for glissile junction, (b) $(111)[01\bar{1}]$ and $(\bar{1}\bar{1}1)[101]$ for Lomer-Cottrell junction, and (c) $(111)[01\bar{1}]$ and $(\bar{1}\bar{1}1)[011]$ for Hirth junction. The vertical and horizontal coordinates are the angles between the junction and the reacting dislocation line directions. The yellow area is the angular space where junction reaction takes place. Outside that region, the interactions result in jog formation.

## 4. Junction reactions in addition to dislocation transport

In our continuum dislocation dynamics model, the reaction terms will be computed numerically during the solution of the transport equations from data provided by separate models of the processes [38]. Hence, whenever and wherever the criteria for junction reactions established in section 3 are satisfied, the coupling term in equation (6) should be activated and the equation should be considered of a transport-reaction type locally. In this section, we show how the local reaction rates are defined.

In the case of collinear annihilation, the two slip systems have the same Burgers vector, which is along the intersection of the two slip planes,

$$\tilde{\mathbf{e}} = \frac{\mathbf{b}^{(1)}}{b} = \frac{\mathbf{b}^{(2)}}{b} \tag{21}$$

When the annihilation criterion (8) is satisfied, the components along the intersection of the two dislocations will annihilate with each other as shown in figure 1. The amount of annihilated dislocation density can be determined based on the fact that dislocation density reduction due to annihilation, $\Delta\rho_{col}^{(1,2)}$, cannot be more than the density itself. This reduction thus satisfies

$$\Delta\rho_{col}^{(1,2)} = \min\ (|\boldsymbol{\rho}^{(1)}\cdot\tilde{\mathbf{e}}|, |\boldsymbol{\rho}^{(2)}\cdot\tilde{\mathbf{e}}|) \tag{22}$$

Here, $|\boldsymbol{\rho}^{(1)}\cdot\tilde{\mathbf{e}}|$ and $|\boldsymbol{\rho}^{(2)}\cdot\tilde{\mathbf{e}}|$ are the density components parallel to the line of intersection of the slip planes and along which the screw component of dislocations falls. Therefore, locally, the dislocation density vectors change over a time step $\Delta t$ according to,

$$\begin{aligned}\boldsymbol{\rho}_{t+\Delta t}^{(1)} &= \boldsymbol{\rho}_t^{(1)} - p\Delta\rho_{col}^{(1,2)}\,\mathrm{sgn}\left(\boldsymbol{\rho}^{(1)}\cdot\tilde{\mathbf{e}}\right)\tilde{\mathbf{e}} \\ \boldsymbol{\rho}_{t+\Delta t}^{(2)} &= \boldsymbol{\rho}_t^{(2)} - p\Delta\rho_{col}^{(1,2)}\,\mathrm{sgn}\left(\boldsymbol{\rho}^{(2)}\cdot\tilde{\mathbf{e}}\right)\tilde{\mathbf{e}}\end{aligned} \tag{23}$$

where $p$ has a value of 1 when the annihilation criterion is satisfied and 0 otherwise. The sign function is used here to ensure that we are subtracting the reacted densities from the original densities, not adding to them, since the direction of $\mathrm{sgn}\left(\boldsymbol{\rho}^{(k)}\cdot\tilde{\mathbf{e}}\right)\tilde{\mathbf{e}}$ always form an acute angle with dislocation $\boldsymbol{\rho}^{(k)}$. To make the density changes by collinear annihilation suitable for use into the evolution equation (6), we express them in a rate form by defining $\dot{\rho}_{col}^{(1,2)} = \Delta\rho_{col}^{(1,2)}/\Delta t$. Hence, the corresponding form of the rate of density change is

$$\begin{aligned}\dot{\boldsymbol{\rho}}_{col}^{(1)} &= -p\dot{\rho}_{col}^{(1,2)}\,\mathrm{sgn}\left(\boldsymbol{\rho}^{(1)}\cdot\tilde{\mathbf{e}}\right)\tilde{\mathbf{e}}, \\ \dot{\boldsymbol{\rho}}_{col}^{(2)} &= -p\dot{\rho}_{col}^{(1,2)}\,\mathrm{sgn}\left(\boldsymbol{\rho}^{(2)}\cdot\tilde{\mathbf{e}}\right)\tilde{\mathbf{e}}.\end{aligned} \tag{24}$$

Thus, we have the collinear annihilation set up in terms of dislocation vector densities. In the current formulation, the rate of collinear annihilation is taken to be the maximum possible rate, annihilating fully the screw component of the smaller density. This implies that, once the dislocations of opposite sign on collinear systems arrive at the same point in space, the annihilation takes place to its fullest. Collinear annihilation is thus treated in exactly the same way opposite dislocations within the same slip system annihilate, which is dictated by our bundle representation of dislocation which requires a mesh resolution on the order of the annihilation distance of opposite dislocations. When the mesh resolution constraint is relaxed, the real local rate of annihilation can be less than this, which can be fixed by a proper statistical analysis of discrete dislocation dynamics results of collinear annihilation.

The rates of glissile junction and sessile junction formation are handled in a similar way. Consider, for example, the glissile junction reaction among dislocations on two different slip planes. The line direction of the junction at the moment it is formed will fall along the intersection of the two slip planes of the reacting dislocations, which is expressed in the form

$$\tilde{\mathbf{e}} = \frac{\mathbf{m}^{(1)} \times \mathbf{m}^{(2)}}{\left|\mathbf{m}^{(1)} \times \mathbf{m}^{(2)}\right|} \tag{25}$$

where $\mathbf{m}^{(1)}$ and $\mathbf{m}^{(2)}$ are unit normal vectors of the two slip planes. If the glissile junction criterion is satisfied, a glissile junction segment $\Delta\boldsymbol{\rho}_{\text{junc}}^{(1,2)} = \Delta\rho_{\text{junc}}^{(1,2)}\tilde{\mathbf{e}}$ or $\Delta\boldsymbol{\rho}_{\text{junc}}^{(1,2)} = -\Delta\rho_{\text{junc}}^{(1,2)}\tilde{\mathbf{e}}$ is formed, with the former being the case when both $\boldsymbol{\rho}^{(1)}\cdot\tilde{\mathbf{e}}$ and $\boldsymbol{\rho}^{(2)}\cdot\tilde{\mathbf{e}}$ are positive and the latter being the case when both $\boldsymbol{\rho}^{(1)}\cdot\tilde{\mathbf{e}}$ and $\boldsymbol{\rho}^{(2)}\cdot\tilde{\mathbf{e}}$ are negative. We remark here that if either $\boldsymbol{\rho}^{(1)}\cdot\tilde{\mathbf{e}}$ or $\boldsymbol{\rho}^{(2)}\cdot\tilde{\mathbf{e}}$ is positive and the other is negative, the energy criterion will not be satisfied and no junction will form. Below we proceed with the form $\Delta\boldsymbol{\rho}_{\text{junc}}^{(1,2)} = \Delta\rho_{\text{junc}}^{(1,2)}\tilde{\mathbf{e}}$, alerting the reader to account for the case $\Delta\boldsymbol{\rho}_{\text{junc}}^{(1,2)} = -\Delta\rho_{\text{junc}}^{(1,2)}\tilde{\mathbf{e}}$ when necessary. The reacting and product dislocation density vectors after the glissile junction is formed have the following forms,

$$\boldsymbol{\rho}_{t+\Delta t}^{(1)} = \boldsymbol{\rho}_t^{(1)} - \Delta\boldsymbol{\rho}_{\text{junc}}^{(1,2)}, \quad \boldsymbol{\rho}_{t+\Delta t}^{(2)} = \boldsymbol{\rho}_t^{(2)} - \Delta\boldsymbol{\rho}_{\text{junc}}^{(1,2)}, \quad \boldsymbol{\rho}_{t+\Delta t}^{(3)} = \boldsymbol{\rho}_t^{(3)} + \Delta\boldsymbol{\rho}_{\text{junc}}^{(1,2)}. \tag{26}$$

In the above, the junction reaction between dislocations with Burgers vector $\mathbf{b}^{(1)}$ and $\mathbf{b}^{(2)}$ lead to the formation of dislocations with Burgers vector $\mathbf{b}^{(3)}$. The direction of $\Delta\boldsymbol{\rho}_{\text{junc}}^{(1,2)}$ is along the intersection of the two slip planes. Dislocations are then subtracted from two parent slip systems and added to the junction slip system according to equation (26). It should be pointed out that, although dislocation density vector on each slip system is changed, the total incompatibility caused by dislocations remains the same. The change of the total dislocation density tensor $\boldsymbol{\alpha}$ is

$$\Delta\boldsymbol{\alpha} = -\Delta\boldsymbol{\rho}_{\text{junc}}^{(1,2)} \otimes \mathbf{b}^{(1)} - \Delta\boldsymbol{\rho}_{\text{junc}}^{(1,2)} \otimes \mathbf{b}^{(2)} + \Delta\boldsymbol{\rho}_{\text{junc}}^{(1,2)} \otimes \mathbf{b}^{(3)} = -\Delta\boldsymbol{\rho}_{\text{junc}}^{(1,2)} \otimes \left(\mathbf{b}^{(1)} + \mathbf{b}^{(2)} - \mathbf{b}^{(3)}\right) = \mathbf{0}. \tag{27}$$

The next question is what is the magnitude of the glissile junction density $\Delta\boldsymbol{\rho}_{\text{junc}}^{(1,2)}$, which is denoted by $\Delta\rho_{\text{junc}}^{(1,2)}$, for given dislocation density vector $\boldsymbol{\rho}^{(1)}$ and $\boldsymbol{\rho}^{(2)}$. Here, a chemical reaction rate equation is adopted [49]. For a reaction $S^{(1)} + S^{(2)} \xrightarrow{c} S^{(3)}$, the reaction rate equation reads

$$\frac{dy_1(t)}{dt} = -cy_1(t)y_2(t), \quad \frac{dy_2(t)}{dt} = -cy_1(t)y_2(t), \quad \frac{dy_3(t)}{dt} = cy_1(t)y_2(t) \tag{28}$$

where $y_1$ and $y_2$ are the concentrations of the reacting species, $y_3$ are the concentration of the product, $c$ is the reaction constant. Equation (28) shows that the effect on the instantaneous rate of change is proportional to the product of the concentrations of the reacting species, see [27,50]. In dislocation junction reactions, the concentrations are replaced by the dislocation density

$$y_1 \to \tilde{\rho}^{(1)} = \boldsymbol{\rho}^{(1)} \cdot \tilde{\mathbf{e}}, \quad y_2 \to \tilde{\rho}^{(2)} = \boldsymbol{\rho}^{(2)} \cdot \tilde{\mathbf{e}} \tag{29}$$

The dislocation density component projected on the intersection of the two slip planes is used. In this case, $c$ represents the junction reaction rate, which should be derived by coarse-graining data from discrete dislocation dynamic simulations, see, for example, see [18,50]. However, since dislocations reactions are governed by their transport and arrangements the junction reaction rate coefficient $c$ must depend on dislocation velocities and dislocation correlations. This statement is elaborated later in the discussion. Finally, based on equations (28) and (29), the amount of glissile junction form, $\Delta \rho_{\text{junc}}^{(1,2)}$, can be derived as

$$\Delta \rho_{\text{junc}}^{(1,2)} = c \tilde{\rho}^{(1)} \tilde{\rho}^{(2)} \Delta t = c(\boldsymbol{\rho}^{(1)} \cdot \tilde{\mathbf{e}})(\boldsymbol{\rho}^{(2)} \cdot \tilde{\mathbf{e}}) \Delta t \tag{30}$$

Dislocations on a slip system can be involved in multiple glissile junction reactions, either as a reactant or a product. The contribution from different junction reactions should be summed. Table 4 shows that, for a FCC crystal, dislocations are possibly involved in 6 glissile junction reactions, 4 as reactant and 2 as product and the total number of types of glissile junction reactions are 24. For glissile junction reaction $S^{(\alpha_k)} + S^{(\beta_k)} \to S^{(\eta_k)}$, we a define junction reaction rate as

$$\mathbf{f}_k(\boldsymbol{\rho}^{(\alpha_k)}, \boldsymbol{\rho}^{(\beta_k)}) = p_k \Delta \boldsymbol{\rho}_{\text{junc}}^{(\alpha_k, \beta_k)} / \Delta t, \quad k = 1, 2, \cdots, 24 \tag{31}$$

where $k$ is used to specify one of the 24 glissile junction reactions and the three involved slip systems are denoted as $\alpha_k, \beta_k$ and $\eta_k$. $p_k$ is an indicator which is 1 if the criterion is satisfied, otherwise 0. This junction reaction rate can be expressed in terms of the reacting dislocation densities as

$$\mathbf{f}_k(\boldsymbol{\rho}^{(\alpha_k)}, \boldsymbol{\rho}^{(\beta_k)}) = p_k c_k (\boldsymbol{\rho}^{(\alpha_k)} \cdot \tilde{\mathbf{e}}_k)(\boldsymbol{\rho}^{(\beta_k)} \cdot \tilde{\mathbf{e}}_k) \tilde{\mathbf{e}}_k, \quad k = 1, 2, \cdots, 24 \tag{32}$$

where $c_k$ is the junction reaction rate coefficient. $\tilde{\mathbf{e}}_k$ is a unit vector along the intersection of slip plane $\alpha_k$ and $\beta_k$ (There will be a negative sign in equation (32) if $\Delta \boldsymbol{\rho}_{\text{junc}}^{(1,2)} = -\Delta \rho_{\text{junc}}^{(1,2)} \tilde{\mathbf{e}}$). According to equation (26), the dislocation densities on the involved three slip systems change with the same amount of dislocations

$$\dot{\rho}_{g\rightarrow}^{(\alpha_k)} = \dot{\rho}_{g\rightarrow}^{(\beta_k)} = \dot{\rho}_{g\leftarrow}^{(\eta_k)} = \mathbf{f}_k(\rho^{(\alpha_k)}, \rho^{(\beta_k)}) \tag{33}$$

where the subscript "$g\rightarrow$" indicates dislocation density is consumed to form glissile junction and "$g\leftarrow$" indicates glissile junction segment is formed on this slip system.

For sessile junctions, equation (32) is applicable but with a different junction reaction rate coefficient $c_k$. Since sessile junction segments do not belong to any of the primary slip system, equation (33) only have terms that consume glide dislocations,

$$\dot{\rho}_{LC\rightarrow}^{(\alpha_r)} = \dot{\rho}_{LC\rightarrow}^{(\beta_r)} = \mathbf{f}_k(\rho^{(\alpha_r)}, \rho^{(\beta_r)}) \quad \text{and} \quad \dot{\rho}_{H\rightarrow}^{(\alpha_s)} = \dot{\rho}_{H\rightarrow}^{(\beta_s)} = \mathbf{f}_k(\rho^{(\alpha_s)}, \rho^{(\beta_s)}) \tag{34}$$

where $r$ and $s$ are used to specify one of the 12 Lomer-Cottrell and Hirth junctions list in table 2 and table 3, respectively. The subscript "$LC\rightarrow$" and "$H\rightarrow$" have similar meaning with "$g\rightarrow$". The consumed densities are stored in local, non-transporting densities representing the LC and H junction densities.

We are now in a position to write the overall transport-reaction equations governing the space-time evolution of all dislocation densities. These equations have the form:

$$\dot{\rho}^{(l)} = \nabla \times (\mathbf{v}^{(l)} \times \rho^{(l)}) + \dot{\rho}_{col}^{(l)} - \sum_{\alpha_k=l} \dot{\rho}_{g\rightarrow}^{(\alpha_k)} + \sum_{\eta_k=l} \dot{\rho}_{g\leftarrow}^{(\eta_k)} - \sum_{\alpha_r=l} \dot{\rho}_{LC\rightarrow}^{(\alpha_r)} - \sum_{\alpha_s=l} \dot{\rho}_{H\rightarrow}^{(\alpha_s)}, \tag{35}$$

where the first term on the right-hand side shows the dislocation evolution due to dislocation transport, and the second term is dislocation collinear annihilation, and the third and fourth terms on show dislocation consumed by forming glissile junction and gained by glissile junction segments formed on this slip system, and the last two terms are dislocation consumed by sessile junctions (Lomer-Cottrell and Hirth junctions). As shown in table 4, there are four glissile junction reactions in which $\rho^{(l)}$ is involved as reactant and two glissile junction reactions in which $\rho^{(l)}$ is involved as product, which means in equation (35), $\dot{\rho}_{g\rightarrow}^{(\alpha_k)}$ has four terms and $\dot{\rho}_{g\leftarrow}^{(\eta_k)}$ has two terms. Similarly, there are two terms for both $\dot{\rho}_{LC\rightarrow}^{(\alpha_r)}$ and $\dot{\rho}_{H\rightarrow}^{(\alpha_s)}$. The most complete form of equation (35) must also include a cross slip rate term, see [38]. By solving equation (35), together with equations (24) (32) (33) and (34), the evolution of the dislocation system with collinear annihilation, glissile junction and sessile junction mechanisms can be studied.

In our simulations, sessile junctions are distinguished based upon their type, and, therefore, the Lomer-Cottrell and Hirth junctions are expressed by density fields, $\rho_{LC}^{(r)}$ and $\rho_{H}^{(s)}$, respectively, throughout the domain of interest. Their evolution is simply expressed as

$$\dot{\boldsymbol{\rho}}_{LC}^{(r)} = \mathbf{f}_k(\boldsymbol{\rho}^{(\alpha_r)}, \boldsymbol{\rho}^{(\beta_r)}), \quad r = 1, 2, \cdots 12, \tag{36}$$

$$\dot{\boldsymbol{\rho}}_{H}^{(s)} = \mathbf{f}_k(\boldsymbol{\rho}^{(\alpha_s)}, \boldsymbol{\rho}^{(\beta_s)}), \quad s = 1, 2, \cdots 12. \tag{37}$$

## 5. Numerical simulations and results

The least squares finite element method with implicit Euler time integration has been used to solve the dislocation transport equations. The numerical scheme can be found in [1]. Within this scheme, the dislocation transport and reactions are treated using operator splitting. In this scheme, the transport equations are solved first at every time step then the density is corrected to account for the reactions. The numerical tests made here are done for FCC crystal structure.

### *5.1. Junction reactions at a material point without dislocation transport*

Before coupling junction reactions with dislocation transport as shown in equation (35), a few tests are performed to show how the dislocation density evolves only by junction reactions. In these tests, uniform dislocation densities are initially assigned to all points for the slip systems of interest. The evolution of dislocations is calculated by equation (35), but only with the junction reaction terms activated while setting the dislocation velocity to be zero. Figure 5 shows two initial dislocation configurations used in subsections 5.1.1 and 5.1.2.

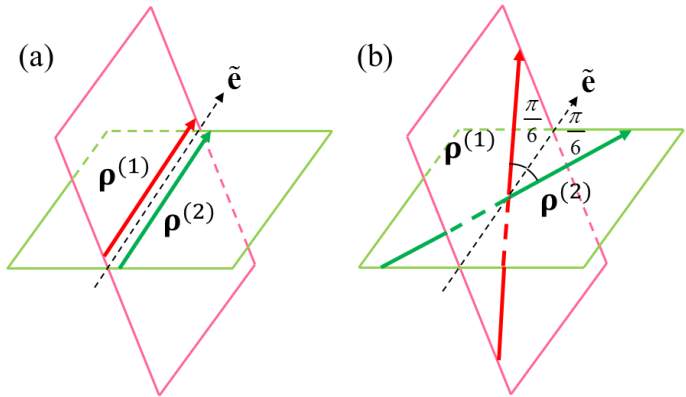

Figure 5. Two initial dislocation configurations prior to junction reactions, (a) parallel dislocations and (b) intersecting dislocations.

### 5.1.1 Junction formed by parallel dislocations

In this test problem, the two reacting dislocation fields, $\rho^{(1)}$ and $\rho^{(2)}$, have line directions parallel to the intersection of the two slip planes, $\tilde{\mathbf{e}}$, as shown in figure 5 (a). The magnitude of the dislocation densities are initially different, with $\rho^{(1)} = 10.0$ μm$^{-2}$ and $\rho^{(2)} = 3.0$ μm$^{-2}$. The junction reaction rate in equation (30) is taken to be $c = 0.1$ μm$^2$ns$^{-1}$. The junction (field) will be formed along $\tilde{\mathbf{e}}$, and it is denoted as $\rho_{LC}$ for Lomer-Cottrell junction and $\rho_H$ for Hirth junction. For Lomer-Cottrell junction, the two reacting slip systems are (111) $[01\bar{1}]$ and $(\bar{1}\bar{1}1)$ [101]. The reaction results in a Lomer-Cottrell junction with $\tilde{\mathbf{e}} = \frac{1}{\sqrt{2}}[1\bar{1}0]$ and $\mathbf{b}_{LC}$ of the type [110]. For Hirth junction, the two reacting slip systems are (111) $[01\bar{1}]$ and $(\bar{1}\bar{1}1)$ [011]. The reaction results in a Hirth junction with $\tilde{\mathbf{e}} = \frac{1}{\sqrt{2}}[1\bar{1}0]$ and $\mathbf{b}_H$ of the type [020]. The evolution of dislocation densities is shown in figure 6 (a) and (b) for Lomer-Cottrell junction and Hirth junction, respectively. As expected, the evolution of dislocation densities in both cases is the same since the reactions rates are the same. The dislocations on the reacting slip systems will be consumed to form the junction. The corresponding density decreases from 10 μm$^{-2}$ to 7 μm$^{-2}$ on slip system 1, and it decreases from 3 μm$^{-2}$ to 0 μm$^{-2}$ on slip system 2, while the same amount is recreated in the form of junction, increasing from 0.0 μm$^{-2}$ to 3.0 μm$^{-2}$. The reaction continues until the smaller density is fully consumed.

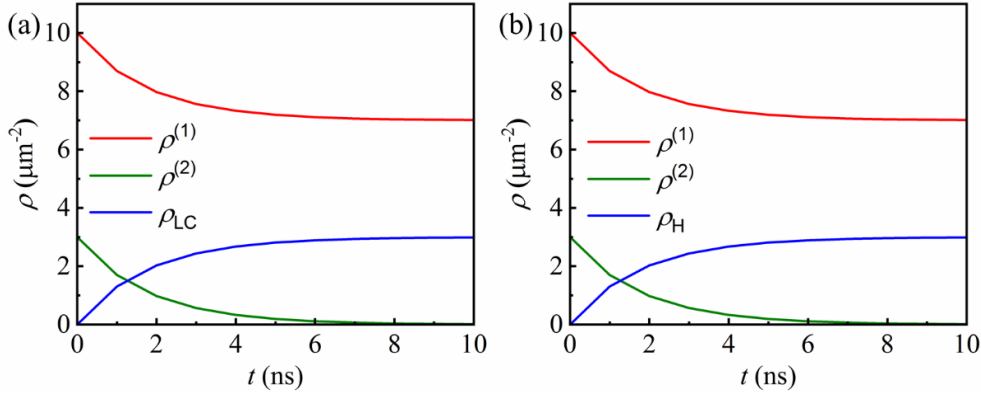

Figure 6. The reactions of two dislocation fields with dislocations parallel to the intersection of the slip planes: (a) evolution of dislocation density for Lomer-Cottrell junction, and (b) evolution of dislocation density for Hirth junction.

### *5.1.2 Junction formed by intersecting dislocations*

The simulation setup in this case is the same as the previous example except that the two reacting dislocations are not parallel. The angle between them and the junction are set to be $\pi/6$ as shown in figure 5 (b). The evolution of dislocation density is shown in figure 7 (a) and (b) for the Lomer-Cottrell and Hirth junction reactions, respectively. One obvious difference from the previous test is that there is no Hirth junction formed at these dislocation orientations, because the energy criterion of forming Hirth junction is not satisfied. So the densities remain the same over time. On the other hand, as Lomer-Cottrell junctions form over a much larger range of orientations (see figure 4), they form in the current example, see figure 7. It is observed, however, that the amount of dislocation density changes is slightly different from the case when the reacting dislocations are initially parallel. For example, the dislocation density on slip system 2 decreases from 3 μm$^{-2}$ to about 1.5 μm$^{-2}$ instead of being fully consumed. Dislocation density on slip system 1 decreases from 10 μm$^{-2}$ to about 7.86 μm$^{-2}$ and the junction density increases from 0 μm$^{-2}$ to 2.6 μm$^{-2}$. The reason why the amount of consumed dislocation is not the same as the junction formed is that junction reaction is performed by vector subtraction while figure 7 only shows the evolution of the scalar densities. Assume that a junction segment $\Delta\boldsymbol{\rho}_{junc}^{(1,2)}$ is formed among slip systems 1 and 2. The dislocation density before and after junction formation on slip system 1 are $\boldsymbol{\rho}^{(1)}$ and $\boldsymbol{\rho}^{(1)} - \Delta\boldsymbol{\rho}_{junc}^{(1,2)}$, and corresponding the scalar dislocation density changes from $|\boldsymbol{\rho}^{(1)}|$ to $|\boldsymbol{\rho}^{(1)} - \Delta\boldsymbol{\rho}_{junc}^{(1,2)}|$. Obviously, the difference is not equal to $|\Delta\boldsymbol{\rho}_{junc}^{(1,2)}|$ unless $\boldsymbol{\rho}^{(1)}$ is parallel to $\Delta\boldsymbol{\rho}_{junc}^{(1,2)}$, which is the case in section 5.1.1. This interpretation physically means that when intersecting dislocations form junctions, they rotate to be parallel and the parallel components begin to form junctions until one of them is completely consumed. Our approach thus accounts for the length change of dislocations due to local rearrangement of the dislocation line configuration when junctions are formed.

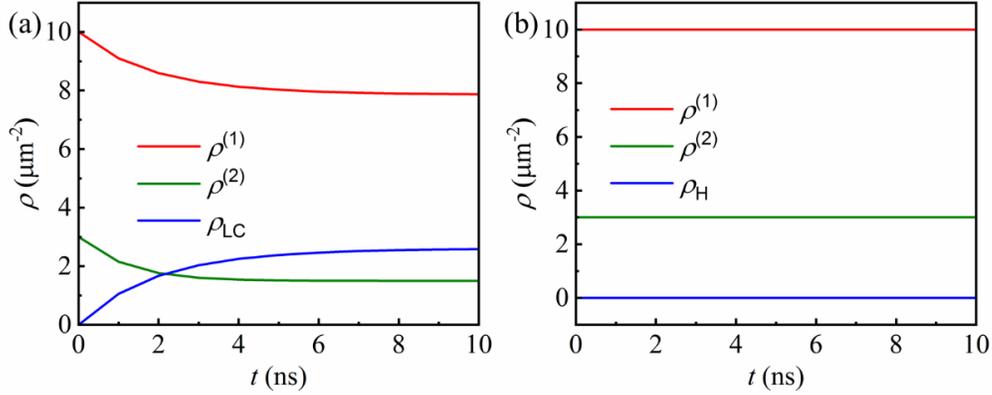

Figure 7. The reaction of two intersecting dislocation densities on two slip systems: (a) evolution of dislocation density for Lomer-Cottrell junctions, and (b) evolution of dislocation density for Hirth junctions.

### *5.1.3 Multiple junction reactions on multiple slip systems*

In this simulation, all of the 12 slip systems of the FCC crystal (table 1) are populated with dislocations. The magnitude and line orientation of the initial (vector) dislocation density are chosen randomly for each slip system. The density is selected in the range 3 $\mu m^{-2}$ to 10 $\mu m^{-2}$. Any of the junction types (glissile, Lomer-Cottrell, and Hirth) can form if the corresponding criteria are satisfied by local orientations of the reacting dislocations. The evolution of dislocation densities, Lomer-Cottrell junction densities and Hirth junction densities are shown in figure 8 (a), (b) and (c), respectively. When reactions take place, the dislocation densities decrease while the junction densities increase. This simulation demonstrates the ability of the model to capture multiple junction reactions simultaneously.

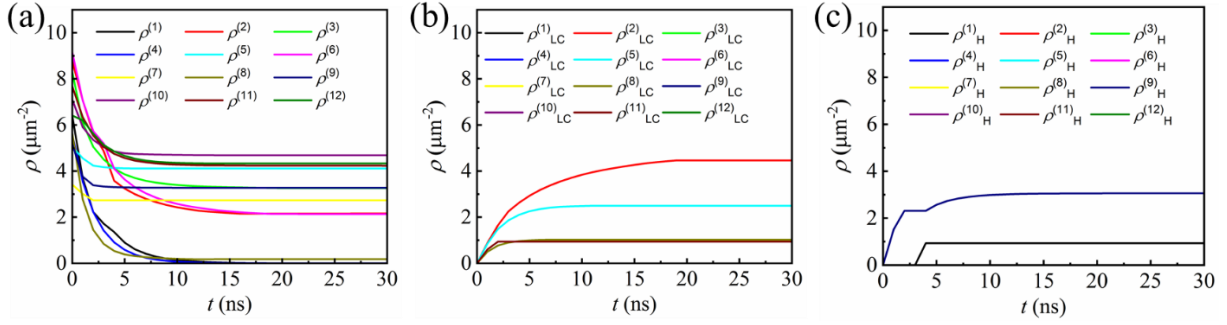

Figure 8. Junction reaction simulation with 12 slip systems of FCC crystal: (a) evolution of dislocation densities, (b) evolution of Lomer-Cottrell junction densities, and (c) evolution of Hirth junction densities.

### 5.2. Junction reactions coupled with dislocation transport

In this section, the dislocation density evolution with both transport and reactions is tested by solving equation (35) using an operator splitting scheme in which the transport problem is solved first then the resulting density is corrected for the reactions at every time step. These tests are designed to reveal the local changes in dislocation configuration due to junction reactions while in motion. A prescribed dislocation velocity is thus chosen for dislocations.

### 5.2.1 Collinear annihilation during two loops expansion

This test shows annihilation of two dislocation loops expanding on two different slip systems. The simulation domain is a $5\mu m \times 5\mu m \times 5.3\mu m$ box, with periodic boundary condition for dislocation evolution,

$$\begin{aligned}
\rho(x=0, y, z) &= \rho(x=L_x, y, z), \\
\rho(x, y=0, z) &= \rho(x, y=L_y, z), \\
\rho(x, y, z=0) &= \rho(x, y, z=L_z).
\end{aligned} \quad (38)$$

The $x$, $y$ and $z$ axes are taken to be along the $[110]$, $[\bar{1}10]$ and $[001]$ crystallographic directions, respectively. The box dimensions are dictated by the mesh used to execute the numerical solution, which is explained in detail in [1]. Slip systems of FCC crystal are used with Burgers vector magnitude 0.256 nm, representative of copper. Two active slip systems are chosen for this test, $(111)[01\bar{1}]$ and $(\bar{1}11)[01\bar{1}]$ ((plane)[slip direction]). On each slip system, there is one dislocation bundle in the form of a loop. The location and line direction of the loops are shown in figure 9 (a) and figure 10 (a). The two dislocation bundles have the same Burgers vector but with opposite directions on the intersection of the two slip planes. Dislocations in these two bundles annihilate with each other when they expand and meet along the line of interaction of the two slip plane. A prescribed dislocation velocity of 0.03 µm/ns is applied on both slip systems. The time step used to solve equation (35) is controlled by a Courant number $C = v_{max}^{(\alpha)} \Delta t / l_{mesh} = 0.45$, where $v_{max}^{(\alpha)}$ is the maximum velocity on slip system $\alpha$ and $l_{mesh} = 62.5$ nm is the mesh size.

Tests without and with collinear annihilation are compared in figure 9 and figure 10, respectively. Different colors are used to represent dislocations on different slip systems. Without collinear annihilation, figure 9, the two loops expand individually. When the dislocations meet on the intersection of the two slip planes, they pass each other as the reaction between them is not activated. On the other hand, with collinear annihilation reaction activaed, figure 10, the two loops interact with each other. At the intersection line of the slip planes, the two dislocations annihilate since they have opposite line directions and the same Burgers vector. The loops start to merge with one another on the line of intersection of the two slip plane to create one continuous bundle of the same Burgers vector extended on the two slip planes. In continuum dislocation dynamics simulations of real systems, only a fraction of the dislocation bundles meeting each other react and become part of the network on the two slip systems. Such situations, therefore, contain aspects of the situations depicted in figure 9 and figure 10.

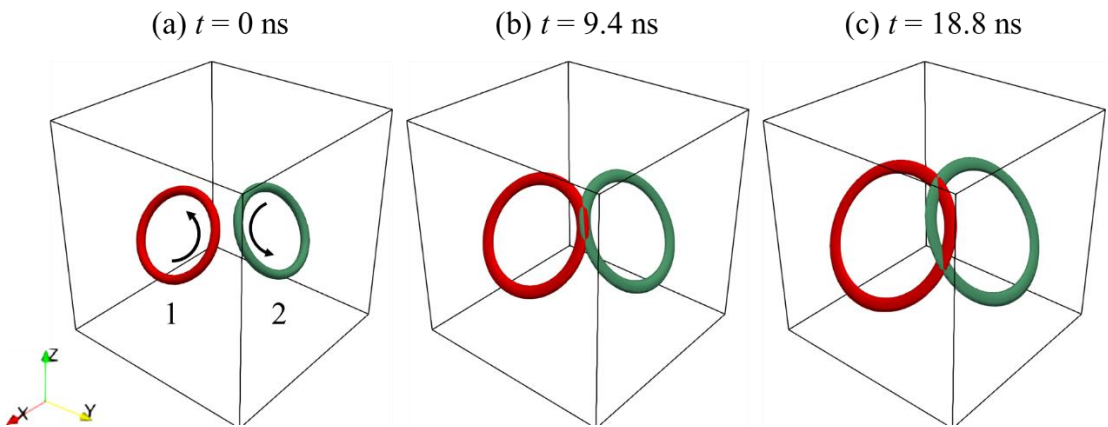

Figure 9. Dislocation bundles bypassing each other at the intersection of their slip planes in the absence of collinear annihilation reaction.

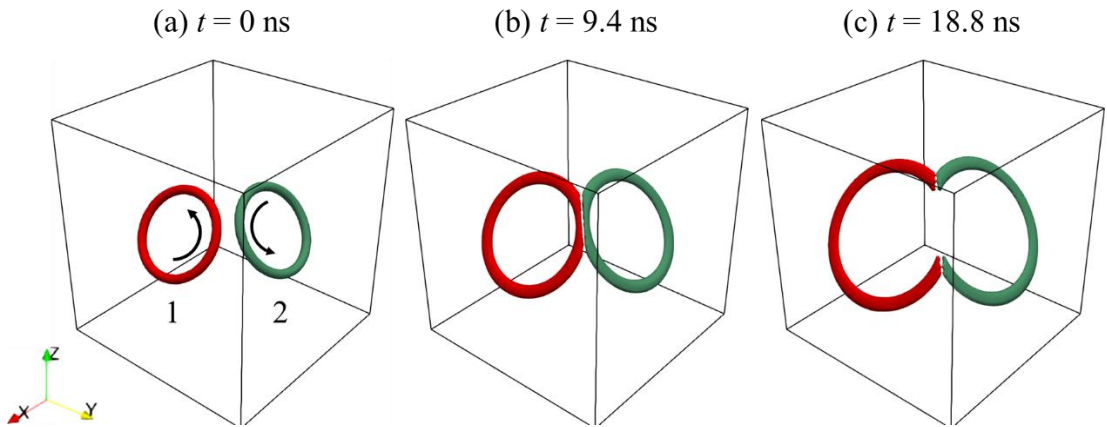

Figure 10. Dislocations on collinear slip systems annihilate each other along the line of interaction of their slip planes and create one continuous bundle of the same Burgers vector extending over the two slip planes.

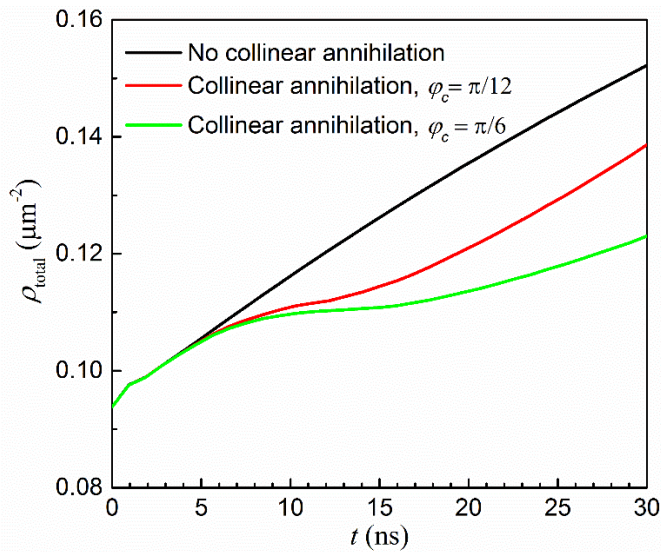

Figure 11. Total dislocation density evolution without collinear annihilation and with collinear annihilation with two different critical angles, $\varphi_c = \pi/12$ and $\varphi_c = \pi/6$.

As show in figure 11, the dislocation density increases linearly without collinear annihilation as there is only the expanding of loops (black line). When collinear annihilation is activated, three stages of the density evolution can be observed (red and green line). Before the two loops encounter each other, the density increases linearly with time following the radius increase. When the loops meet at the intersection, the density increase slows down due annihilation. After all dislocations satisfying the collinear annihilation criteria have annihilated, loop expansion dominates the density evolution again (red line). The latter stage is sensitive to the critical angle for collinear annihilation as depicted by the green line, see equation (8). It should be pointed out that the 'knee' in the solution at $t \approx 1$ ns is due to the numerical smoothing of the difference between the analytically generated initial dislocation density and its version interpolated by the shape function in the finite element scheme.

### *5.2.2 Glissile junction formed during two loops expansion*

Next, we test glissile junction formation among expanding loops. A similar dislocation configuration as that used with the collinear annihilation test is used, but the slip systems and the line direction of the loops are different. Again, the simulation domain is a $5\mu m \times 5\mu m \times 5.3\mu m$ box, with periodic boundary condition for dislocation evolution but three slip systems are now active. Two reacting slip systems are chosen as (111) $[01\bar{1}]$ and $(\bar{1}11)$ [101], while the glissile junction will be formed on slip system $(\bar{1}11)$ [110]. The slip plane of the glissile junction is the same as slip system 2 and its Burgers vector is given by the sum of the reacting Burgers vectors, $\mathbf{b}^{(3)} = \mathbf{b}^{(1)} + \mathbf{b}^{(2)}$. Initially, there is one dislocation loop on each of the slip systems as shown in figure 12 (a) and figure 13 (a). A prescribed dislocation velocity of 0.03 μm/ns is applied on all slip systems. The time step used is chosen as in the previous test.

Tests without and with glissile junction reactions are compared in figure 12 and figure 13, respectively. Different colors are used to represent dislocations on different slip systems, red and green for the two reacting slip systems and blue for the glissile junction formed on the third slip system. When no glissile junction reaction is allowed and only dislocation transport is considered, the dislocation loops expand on their respective slip plane. In this case, no dislocations form on slip system 3. On the other hand, when the two loops react with with one another at the intersection of the two slip planes, dislocations are consumed on the two reacting slip systems and dislocation segments emerge on slip system 3 at the same location (figure 13). As the newly formed dislocations are mobile, they will bow out and expand just like Frank-Read source but with two moving end points (figure 13). The final result is that the dislcoations on the three slip systems appear connected at the triple joint points on the line of intersection of the reacting slip planes.

This phenomenon have been shown in earlier discrete dislocation dynamics [5] and it represents an effective dislocation multiplication mechanism.

The evolution of the total dislocation density is shown in figure 14. Parts (a) and (b) of the figure show the densities on the reacting slip systems. Part (c) shows the densities on the glissile junction slip system. As discussed in section 5.2.1, dislocation density evolution is controlled by both dislocation expansion and dislocation reactions. The dislocation density on the two reacting slip systems increases linearly without glissile junction and no dislocation on the third slip system. During glissile junction reaction, however, the density evolution on the two reacting slip systems shows three stages, since the glissile junction formation consumes dislocations on the reacting slip systems, which slows down the increase of dislocation densities. In the meantime, the dislocation density on the third slip system begins to increase due to the newly formed glissile junction and the expansion of the glissile junction. The rate of glissile junction reaction has been varied for comparison, and the results show that a larger glissile junction reaction rate consumes more dislocations on the reacting slip systems leading to higher density of the product species.

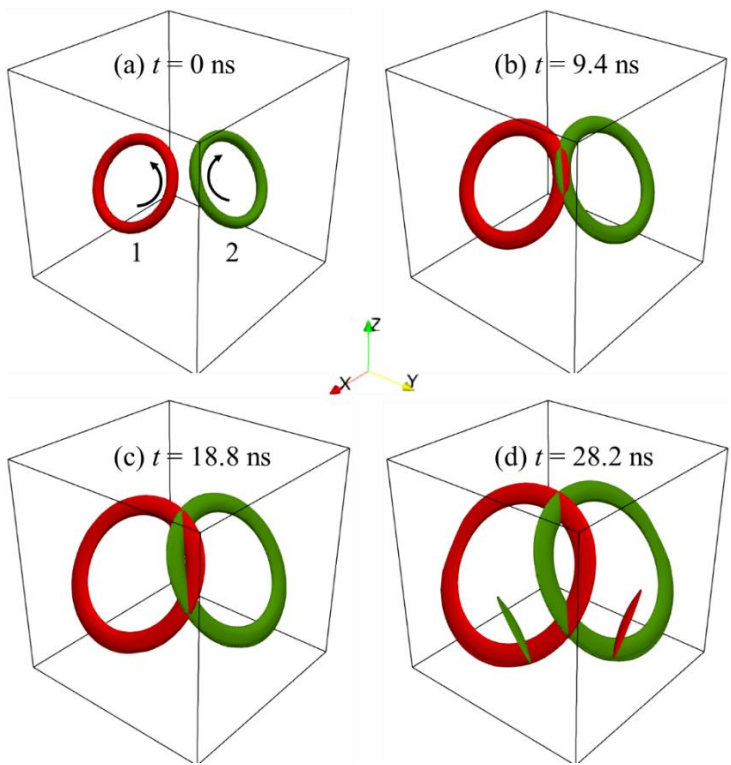

Figure 12. Evolution of the two-loop dislocation configuration without glissile junction reaction.

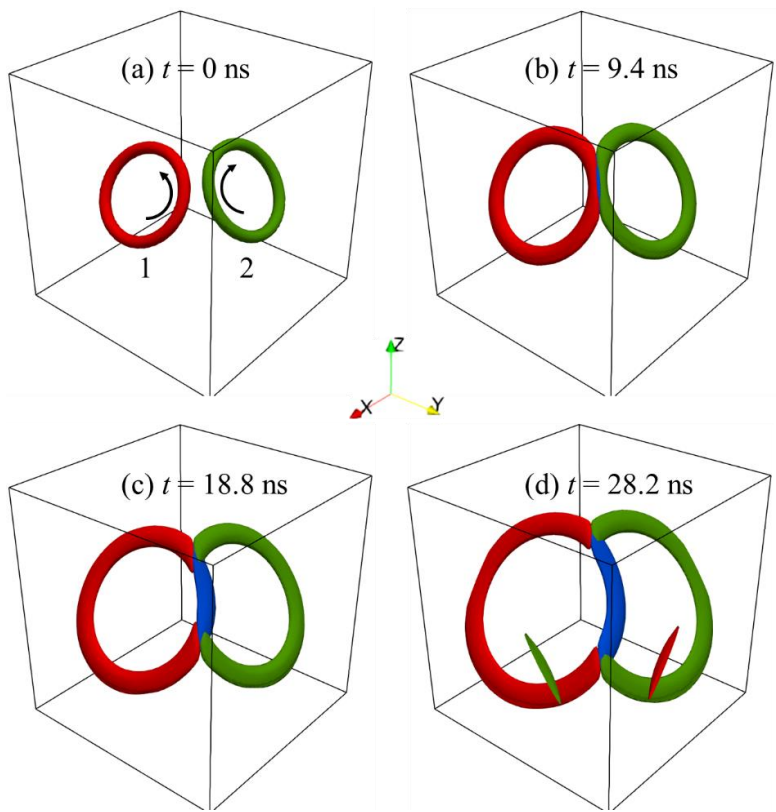

Figure 13. Evolution of the two-loop dislocation configuration with glissile junction formation. The formed junction provides closure to the open dislocation loops on the reacting slip systems.

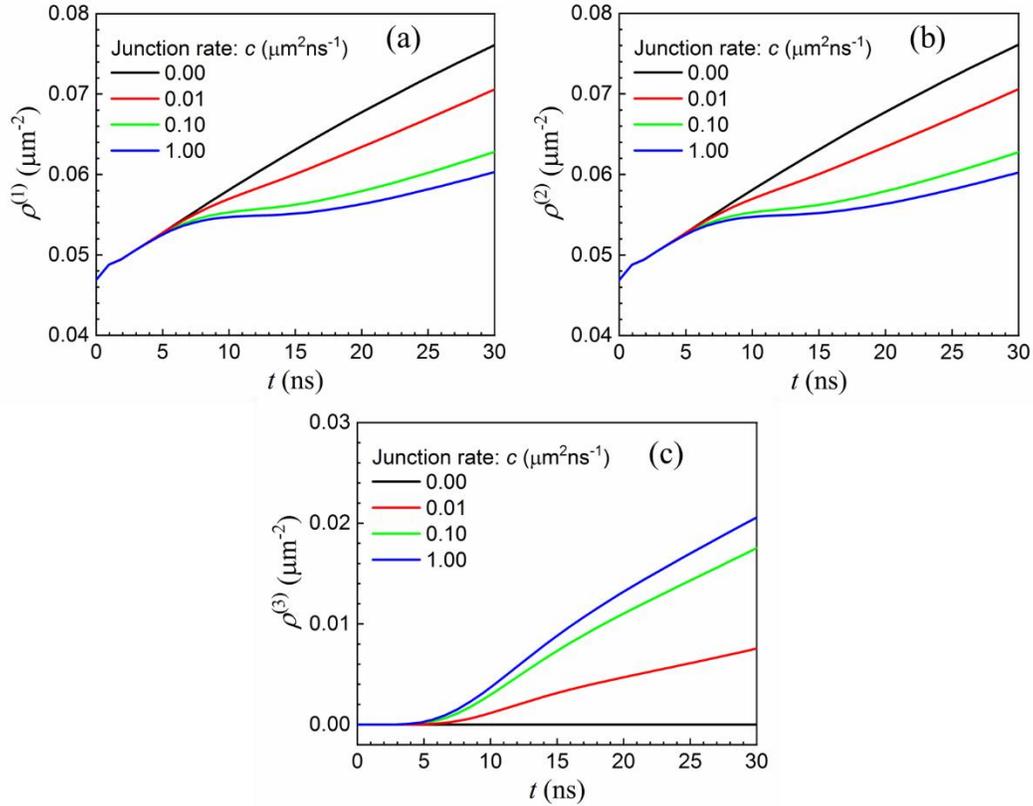

Figure 14. Dislocation density evolutions in glissile junction test. Parts (a) and (b) show the density on the reacting slip systems, respectively. Part (c) shows the density on the glissile junction slip system.

### *5.2.3 Sessile junction formed during two loops expansion*

Sessile junction formation is tested next. Junctions of Lomer-Cottrell type are chosen as an example, while Hirth junction formation tests yielded similar results. The simulation domain and boundary condition used in the previous tests are considered in the current test. To form a Lomer-Cottrell junction, two reacting slip systems, (111) [01$\bar{1}$] and ($\bar{1}\bar{1}$1) [101], are considered. These slip systems are populated with dislocation loop bundles expanding at a constant velocity of 0.03 μm/ns as shown in figure 15 (a). The evolving dislocation structure is shown in figure 15 at different time steps. When the two loops reach each other at the intersection of the two slip planes, the reacting dislocations are consumed and the junction is formed along the line of intersection of the slip planes of the former. Unlike the glissile junction case, which form a Frank-Read like source, the Lomer-Cottrell junction is not mobile and it remains at the intersection of the slip planes with a length increasing as the reacting loops expand.

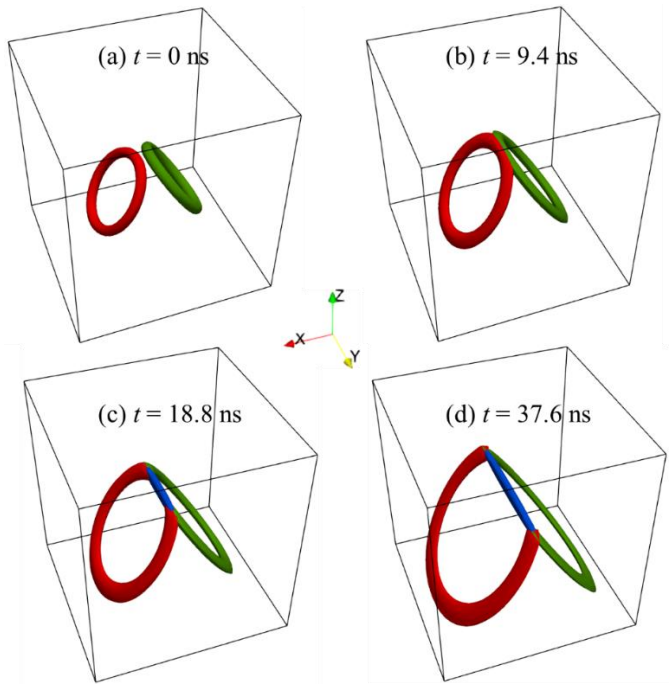

Figure 15. Lomer-Cottrell junction formation during two loops expansion. The junction is straight along the interaction of the two slip planes.

### *5.2.4 Multiple junction reactions with multiple slip systems*

Junction reactions among dislocations on all 12 slip systems of an FCC crystal are also tested with dislocation transport. An initial dislocation structure consisting of four loops per slip systems is formed for this purpose, see figure 16. The centers of these loops are randomly placed in the crystal with radii ranging from 1 µm to 2 µm. A prescribed dislocation velocity of 0.03 µm/ns is chosen for all dislocation loops to make them expand. The dislocation density at the end of the simulation (t = 187.5 ns) is shown in figure 17. It can be seen that, and as should be, the Lomer-Cottrell junction and Hirth junctions always take place at the intersection of the slip planes. In this simulation, Lomer-Cottrell junction density is higher than Hirth junction density. Since the same rate parameter for junction formation is used in both Lomer-Cottrell and Hirth junctions and all slip systems are active with the same dislocation velocity, the differences in frequency of occurrence of both types of junctions can only be partly attributed to the satisfaction of the energy criterion (see figure 4). However, as the velocity of dislocations is prescribed, as opposed determined in terms of the local stress via a mobility law, it must be pointed out that the microstructure observed, including the junctions formed, is not the same as that which would be observed in crystals undergoing deformation under external loads.

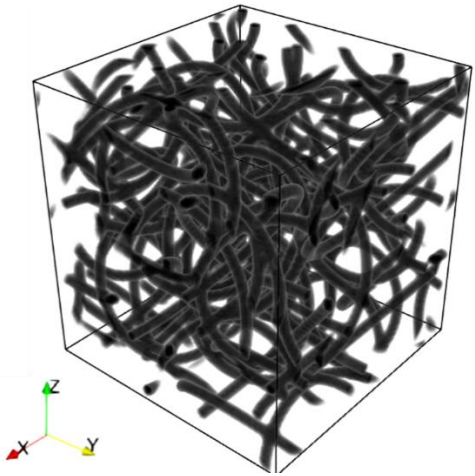

Figure 16. Initial dislocation structure for the case of transport and reactions among all slip systems.

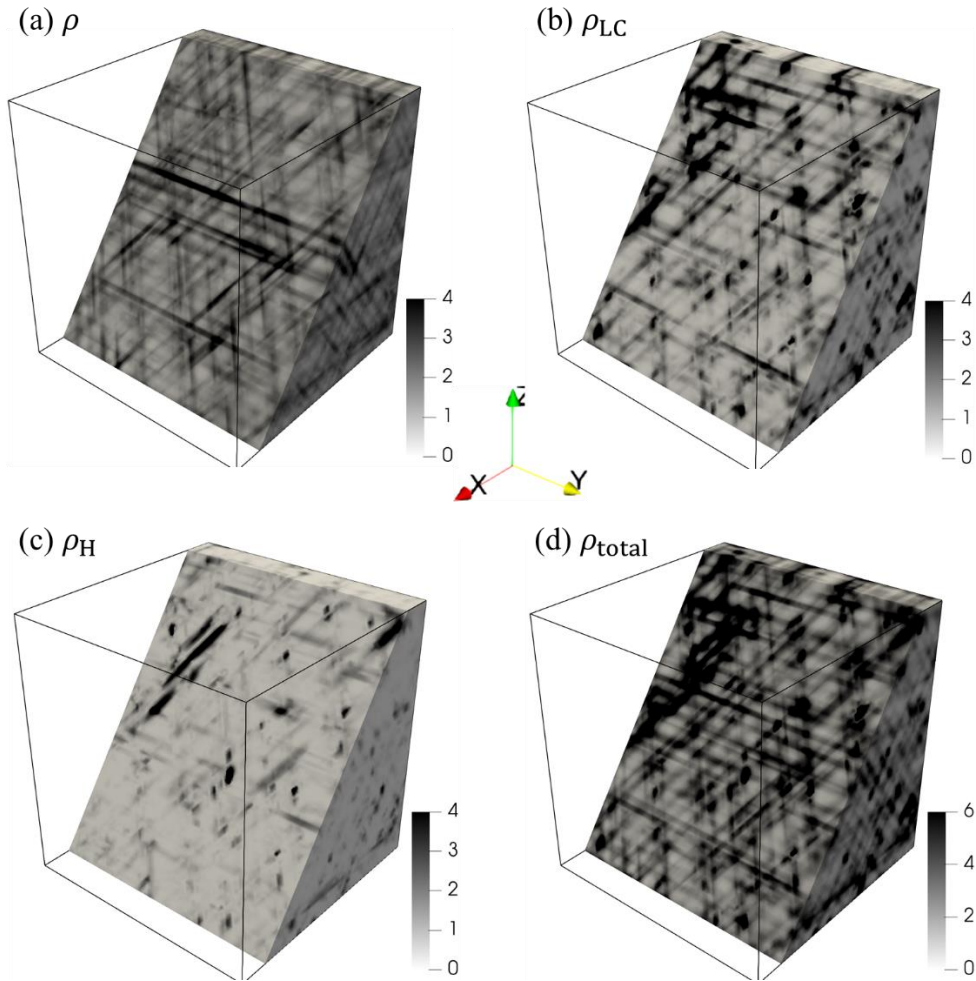

Figure 17. The dislocation density at the end of the simulation. (a) Glide dislocation density $\rho$, (b) Lomer-Cottrell junction density $\rho_{LC}$, (c) Hirth junction density $\rho_H$, (d) Total dislocation density $\rho_{total} = \rho + \rho_{LC} + \rho_H$.

## 6. Discussion

In the previous sections, we primarily focused on the numerical implementation of dislocation reactions in the vector density-based continuum dislocation dynamics formalism, where the dislocation line orientation was included in the criteria determining whether a specific type of dislocation reaction is to take place (Eqs. (8) and (18)). A consequence of the linear nature of dislocations is that dislocations remain connected lines after reaction, as the Frank's rule should be satisfied. In a continuum description of dislocations, the dislocation line connectivity requirement is expressed by the condition

$$\nabla \cdot \boldsymbol{\alpha} = \mathbf{0}, \qquad (39)$$

with $\boldsymbol{\alpha} = \sum_{l} \boldsymbol{\rho}^{(l)} \otimes \mathbf{b}^{(l)}$ being the total dislocation density tensor. The reactions formulated in Eq. (24) (collinear annihilation), Eq. (33) (glissile junction), and Eq. (34) (sessile junction) all comply with this rule. However, they are not the only possible ways to comply this rule. In the following, other forms will be discussed. Consider two reacting slip systems with Burgers vectors $\mathbf{b}^{(1)}$ and $\mathbf{b}^{(2)}$, and assume that the resulting dislocation (junction) has Burgers vector $\mathbf{b}^{(3)}$ such that $\mathbf{b}^{(3)} = \mathbf{b}^{(1)} + \mathbf{b}^{(2)}$. (For the collinear annihilation case, $\mathbf{b}^{(1)} = -\mathbf{b}^{(2)}$, and $\mathbf{b}^{(3)} = \mathbf{0}$). If $\tilde{\mathbf{e}}$ is a unit vector along the intersection of the two reacting slip planes, a general reaction formula satisfying Eq. (39) can be written in the form

$$\begin{aligned}
\boldsymbol{\rho}^{(1)}_{t+\Delta t} &= \boldsymbol{\rho}^{(1)}_{t} - \Delta\rho^{(1,2)}_{\text{junc}} \tilde{\mathbf{e}} \\
\boldsymbol{\rho}^{(2)}_{t+\Delta t} &= \boldsymbol{\rho}^{(2)}_{t} - \Delta\rho^{(1,2)}_{\text{junc}} \tilde{\mathbf{e}} \\
\boldsymbol{\rho}^{(3)}_{t+\Delta t} &= \boldsymbol{\rho}^{(3)}_{t} + \Delta\rho^{(1,2)}_{\text{junc}} \tilde{\mathbf{e}}
\end{aligned} \qquad (40)$$

in which $\Delta\rho^{(1,2)}_{\text{junc}}$ ($\Delta\rho^{(1,2)}_{\text{col}}$) is the scalar amount of junction formed (collinear annihilation) during time increment $\Delta t$. In section 4, $\Delta\rho^{(1,2)}_{\text{junc}}$ was expressed as a function of $\boldsymbol{\rho}^{(1)}$ and $\boldsymbol{\rho}^{(2)}$, but was handled differently in collinear annihilation and in junctions. Specifically, in the collinear annihilation case (Eq. (22)), $\Delta\rho^{(1,2)}_{\text{col}}$ was not dependent on the time increment $\Delta t$ or a rate coefficient $c$, while it does in the junction case (Eq. (30)). A question that poses itself here is: Can we use a similar form for collinear annihilation rate as that used for junctions? The answer is positive. In fact, the following equation can be used to replace Eq. (22):

$$\Delta\rho^{(1,2)}_{\text{col}} = c \, | \boldsymbol{\rho}^{(1)} \cdot \tilde{\mathbf{e}} | \, | \boldsymbol{\rho}^{(2)} \cdot \tilde{\mathbf{e}} | \, \Delta t . \qquad (41)$$

In contrast with Eq. (22), in using Eq. (41) not all dislocations will be consumed in a given time $\Delta t$, and the portion of the dislocation density involved in the annihilation depends on the rate coefficient $c$. The reason for using Eq. (22) in section 4 was that we assumed that collinear annihilation happens much fast than dislocation transport, which is dictated by the spatial resolution constraint explained earlier. As such, the use of a suitable rate coefficient in Eq. (41) would make the form equivalent to (22). On the other hand, the junction reaction formula (Eq. (30)) is established in spirit of chemical reaction. However, care must be taken since dislocations have a linear nature and that, for them to react, they must reach each other via transport and hence the rate of reaction must be proportional to the rate at which dislocation reach each other. In this sense, the dislocation velocity is important.

In the literature, the rate forms of dislocation reaction vary in CDD models [5,32,36,37]. To include the observed dislocation source mechanism in their DDD simulations, Stricker and Weygand [5] proposed a junction density formation rate based on the collision frequency:

$$\dot{\rho}_{junc}^{(3)} = c_1\sqrt{\rho^{(2)}}\rho^{(1)}v^{(1)} + c_2\sqrt{\rho^{(1)}}\rho^{(2)}v^{(2)} \qquad (42)$$

where $c_1$ and $c_2$ are constants, which depend on the effective junction length generated by the individual junction process, $\rho^{(1)}$ and $\rho^{(2)}$ are the reacting species, with velocities $v^{(1)}$ and $v^{(2)}$, respectively, and $\rho_{junc}^{(3)}$ is the product dislocation species. In the above equation, the junction rate is proportional to the dislocation velocity. This model is further elaborated by Monavari and Zaiser [36] and Sudmanns et al. [37] in their CDD models. In the latter models, the length of the junction is assumed to be proportional to the average distance between junction points, which scales with the mean dislocation spacing $1/\sqrt{\rho}$. A similar idea can be used in our model. However, in order to remain consistent with our vector-density representation of dislocations, not only the scalar dislocation density but also the line orientation should be considered in determining the collision frequency. For a given material point with dislocation density vectors $\boldsymbol{\rho}^{(1)}$ and $\boldsymbol{\rho}^{(2)}$, with corresponding dislocation velocities $\mathbf{v}^{(1)}$ and $\mathbf{v}^{(2)}$, the following form has been derived to calculate the number of collision (new encounters) per unit volume between $\boldsymbol{\rho}^{(1)}$ and $\boldsymbol{\rho}^{(2)}$ during time $\Delta t$:

$$N_{junc} = \left|(\mathbf{v}^{(2)} - \mathbf{v}^{(1)})\cdot\boldsymbol{\rho}^{(1)}\times\boldsymbol{\rho}^{(2)}\right|\Delta t. \qquad (43)$$

The rate of encounter of reacting dislocations is thus given by the magnitude of the triple scalar product of the relative dislocation velocity $(\mathbf{v}^{(2)} - \mathbf{v}^{(1)})$ and the density vectors $\boldsymbol{\rho}^{(1)}$ and $\boldsymbol{\rho}^{(2)}$ of the dislocations involved in the reaction. The derivation of Eq. (43) is discussed in details in the Appendix. If the length of the junction is $l_{junc}^{(1,2)}$, Eq. (30) for the junction density increment can then be replaced with

$$\Delta\rho_{junc}^{(1,2)} = \left|(\mathbf{v}^{(2)} - \mathbf{v}^{(1)})\cdot\boldsymbol{\rho}^{(1)}\times\boldsymbol{\rho}^{(2)}\right|l_{junc}^{(1,2)}\Delta t. \qquad (44)$$

It should be pointed out that $l_{\text{junc}}^{(1,2)}$ is not a constant; it is rather a function of the orientation of the dislocation densities involved, i.e., $l_{\text{junc}}^{(1,2)}(\varphi^{(1)}, \varphi^{(2)})$, and it can be determined by either DDD simulation [6] or analytical solution based on energy, but the exact form is left for a future publication.

Another important factor influencing dislocation reactions is dislocation correlation [20,51]. Since the density representation of dislocations in CDD is a coarse-grained measure of discrete dislocation lines, Eq. (44) should include additional correlation terms from statistics. The correlations can be derived from DDD simulations [28]. This part is, however, out scope of the current paper, the purpose of which is to incorporate, both theoretically and numerically, dislocation reactions in the vector density-based CDD framework. No matter which formula is used (Eq. (22) or Eq. (41), Eq. (30) or Eq. (44)), our numerical treatment of the reaction terms within the CDD equations will not change.

## 7. Closing remarks

A continuum dislocation dynamics model [1] has been extended to consider dislocation junction reactions among dislocations. In this model, the dislocation densities on various slips systems are represented by vector fields, one per slip system, with the individual density fields having unique dislocation line direction at each point in space. Such a detailed picture of the dislocation systems enabled us to build all types of dislocation reactions into the coupled set of transport equations governing the evolution of the dislocation systems. The reaction terms on these equations represent collinear annihilation, glissile junction formation, and sessile junction formation. The latter includes Lomer-Cottrell junction and Hirth junction types. The annihilation reactions couple the density evolution on two slip systems at a time, while the junction reactions couple the densities on three slip systems at a time.

A rigorous formulation of the reaction rates in terms of the vector density fields has been established in sections 3 and 4. In doing so, an energy-based criterion for junction formation originally cast in terms of line directions and Burgers vectors of the reacting dislocations has been generalized to the continuum density representation setting, which is applicable to all kinds of junction reactions. This criterion was formulated such that the local energy will become smaller upon the formation of the junction. Consideration of the line directions in addition to Burgers vectors of the reacting dislocations enables us to demarcate the short-range encounters of dislocations that lead to junction formation or jog creation in a continuum dislocation density representation setting.

The quasi-chemical rate form of various types of reactions is formulated in terms of the products of the reaction densities, with phenomenological rate parameters, see section 4. In these rate forms, the rate coefficients are assumed to be fixable by some other type modeling, e.g., by statistical analysis of the equivalent discrete dislocation system. An example of such type of models can be found in [37,50]. The rates of reactions are found so as to preserve the local Burgers vector of the reacting slip systems, pair-wise for collinear annihilation of triplet-wise for the junction reactions.

A finite element implementation of the transport-reaction equations has been carried out, and several tests have been performed to investigate the local network changes by annihilation or junction reactions. The model has been tested for FCC crystal structure. The tests, although showing intuitive results, they demonstrate the possibility of representing dislocation reactions systematically in a continuum dislocation dynamics framework. The remarkable aspects of these results are the Frank-Read like behavior of glissile junctions with moving endpoint and the expansion the sessile junctions of Lomer-Cottrell and Hirth type along the intersection of the two slip planes of the reacting dislocations. While these scenarios have been previously reproduced in discrete dislocation dynamics models, e.g., in [5], and it is demonstrated here that they can also be reproduced by continuum dislocation dynamics as well.

In all of the tests presented here, the dislocation velocity was prescribed and the focus has been on solving the transport-reaction equations to demonstrate reactions in continuum dislocation dynamics. Coupling with crystal mechanics will enable a self-consistent solution of the mesoscale plasticity problem so as to determine the impact of dislocation reactions on the dislocation patterning and stress-strain behavior. This is the subject of a future communication.

## Appendix

In this appendix, the rate of encounters of two dislocation densities $\rho^{(1)}$ and $\rho^{(2)}$ at a point in the crystal is formulated. In doing so, we consider a crystal element as shown in figure a1. In parts (b) and (c) of the figure, the red and green arrows represent the two dislocation densities that overlap in space over the element shown in part (a). The volume element is selected so that the $x$ axis is along the direction of $\boldsymbol{\rho}^{(1)}$ with length $l_x$, $y$ axis along the direction of $\boldsymbol{\rho}^{(2)}$ with length $l_y$, and $z$ axis along the direction of the relative velocity, $(\mathbf{v}^{(2)} - \mathbf{v}^{(1)})$, with length $l_z$. The volume $V$ of this element is given by the triple scalar product

$$V = \left| l_z \frac{(\mathbf{v}^{(2)} - \mathbf{v}^{(1)})}{\|\mathbf{v}^{(2)} - \mathbf{v}^{(1)}\|} \cdot l_x \frac{\boldsymbol{\rho}^{(1)}}{\|\boldsymbol{\rho}^{(1)}\|} \times l_y \frac{\boldsymbol{\rho}^{(2)}}{\|\boldsymbol{\rho}^{(2)}\|} \right| = l_x l_y l_z \frac{\left|(\mathbf{v}^{(2)} - \mathbf{v}^{(1)}) \cdot \boldsymbol{\rho}^{(1)} \times \boldsymbol{\rho}^{(2)}\right|}{\|\mathbf{v}^{(2)} - \mathbf{v}^{(1)}\| \cdot \|\boldsymbol{\rho}^{(1)}\| \cdot \|\boldsymbol{\rho}^{(2)}\|}, \quad (A1)$$

with $\|\cdot\|$ referring to the norm of a vector and $|\cdot|$ to the absolute value of a scalar. Here we assume that the dislocations are locally distributed in a uniform fashion in the volume element. Let us now consider the number of dislocation intersections of faces $y$-$z$ and $x$-$z$ in parts (b) and (c) of figure a1. By definition of dislocation density, these number are

$$N^{(1)} = \frac{\|\boldsymbol{\rho}^{(1)}\| V}{l_x} \quad \text{and} \quad N^{(2)} = \frac{\|\boldsymbol{\rho}^{(2)}\| V}{l_y}, \quad (A2)$$

respectively. For a given time increment $\Delta t$, the displacement of $\boldsymbol{\rho}^{(2)}$ relative to $\boldsymbol{\rho}^{(1)}$ is $(\mathbf{v}^{(2)} - \mathbf{v}^{(1)})\Delta t$. Then for each dislocation in $\boldsymbol{\rho}^{(2)}$, it will intersect with $N^{(1)} \frac{1}{l_z} \|\mathbf{v}^{(2)} - \mathbf{v}^{(1)}\| \Delta t$ dislocations of the type $\boldsymbol{\rho}^{(1)}$. So the total number of new encounters (collisions), $N_{\text{junc}}$, between the densities $\boldsymbol{\rho}^{(1)}$ and $\boldsymbol{\rho}^{(2)}$ per unit volume is

$$N_{\text{junc}} = \frac{1}{V} N^{(1)} N^{(2)} \frac{1}{l_z} \|\mathbf{v}^{(2)} - \mathbf{v}^{(1)}\| \Delta t = \left|(\mathbf{v}^{(2)} - \mathbf{v}^{(1)}) \cdot \boldsymbol{\rho}^{(1)} \times \boldsymbol{\rho}^{(2)}\right| \Delta t. \quad (A3)$$

The variable $N_{\text{junc}}$ defined above represents the number density of junctions per unit volume forming over time $\Delta t$. This number when weighted by the average length of junctions at the junction birth would be the junction density measured in dislocation line length, $\Delta \rho_{\text{junc}}^{(1,2)}$, which is given in Eq. (44).

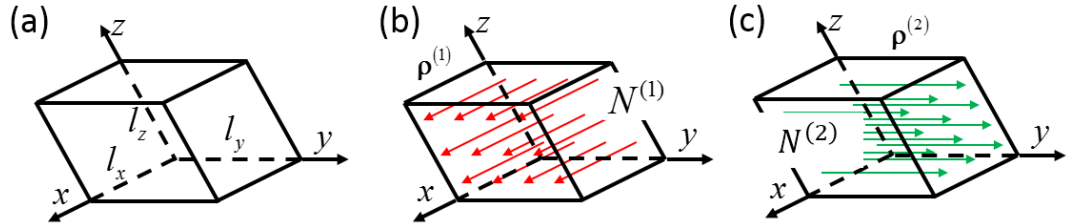

Figure A1 A volume element used to calculate the number of collisions during dislocation reaction.

## Acknowledgement

This work is supported by the US Department of Energy, Office of Science, Division of Materials Sciences and Engineering, through award number DE-SC0017718 at Purdue University. The authors are grateful to the referees for an excellent technical feedback on the original manuscript.

## References

[1]     Xia S and El-Azab A 2015 Computational modelling of mesoscale dislocation patterning and plastic deformation of single crystals *Model. Simul. Mater. Sci. Eng.* **23** 055009

[2]     Taylor G I 1934 The Mechanism of Plastic Deformation of Crystals. Part I. Theoretical *Proc. R. Soc. A Math. Phys. Eng. Sci.* **145** 362–87

[3]     Hirsch P B, Horne R W and Whelan M J 1956 LXVIII. Direct observations of the arrangement and motion of dislocations in aluminium *Philos. Mag.*

[4]     Sills R B, Bertin N, Aghaei A and Cai W 2018 Dislocation Networks and the Microstructural Origin of Strain Hardening *Phys. Rev. Lett.* **121** 85501

[5]     Stricker M and Weygand D 2015 Dislocation multiplication mechanisms - Glissile junctions and their role on the plastic deformation at the microscale *Acta Mater.* **99** 130–9

[6]     Weygand D, Friedman L H, Giessen E Van der and Needleman A 2002 Aspects of boundary-value problem solutions with three-dimensional dislocation dynamics *Model. Simul. Mater. Sci. Eng.* **10** 306

[7]     Arsenlis A, Cai W, Tang M, Rhee M, Oppelstrup T, Hommes G, Pierce T G and Bulatov V V 2007 Enabling strain hardening simulations with dislocation dynamics *Model. Simul. Mater. Sci. Eng.* **15** 553–95

[8]     Po G, Mohamed M S, Crosby T, Erel C, El-Azab A and Ghoniem N 2014 Recent Progress in Discrete Dislocation Dynamics and Its Applications to Micro Plasticity *Jom* **66** 2108–20

[9]     Evers L P, Brekelmans W A M and Geers M G D 2004 Non-local crystal plasticity model with intrinsic SSD and GND effects *J. Mech. Phys. Solids* **52** 2379–401

[10]    Lin P, Liu Z, Cui Y and Zhuang Z 2015 A stochastic crystal plasticity model with size-dependent and intermittent strain bursts characteristics at micron scale *Int. J. Solids Struct.* **69–70** 267–76

[11]    Lin P, Liu Z and Zhuang Z 2016 Numerical study of the size-dependent deformation morphology in micropillar compressions by a dislocation-based crystal plasticity model *Int. J. Plast.* **87** 32–47

[12]    Lin P, Nie J F, Liu Z L and Zhuang Z 2017 Study of two hardening mechanism caused by geometrically necessary dislocations in thin films with passivation layer *To Be Submitt.*

[13]    Franciosi P, Berveiller M and Zaoui A 1980 Latent hardening in copper and aluminium single crystals *Acta Metall.* **28** 273–83

[14]    Kubin L, Devincre B and Hoc T 2008 Toward a physical model for strain hardening in fcc crystals *Mater. Sci. Eng. A* **483–484** 19–24

[15]    Madec R, Devincre B, Kubin L, Hoc T and Rodney D 2003 The role of collinear interaction in dislocation-induced hardening *Science (80-. ).* **301** 1879–82

[16]    Devincre B, Kubin L and Hoc T 2006 Physical analyses of crystal plasticity by DD simulations *Scr. Mater.* **54** 741–6

[17]    Mishra A and Alankar A 2019 Revisiting dislocation reactions and their role in uniaxial deformation of copper single crystal micro-pillars *Model. Simul. Mater. Sci. Eng.* **27** 055010

[18]    Roters F, Diehl M, Shanthraj P, Eisenlohr P, Reuber C, Wong S L, Maiti T, Ebrahimi A, Hochrainer T, Fabritius H-O, Nikolov S, Friák M, Fujita N, Grilli N, Janssens K G F, Jia N, Kok P J J, Ma D, Meier F, Werner E, Stricker M, Weygand D and Raabe D 2019 DAMASK – The Düsseldorf Advanced Material Simulation Kit for modeling multi-

physics crystal plasticity, thermal, and damage phenomena from the single crystal up to the component scale *Comput. Mater. Sci.* **158** 420–78

[19] Zaiser M, Miguel M-C and Groma I 2001 Statistical dynamics of dislocation systems: The influence of dislocation-dislocation correlations *Phys. Rev. B* **64** 224102

[20] Groma I, Csikor F F and Zaiser M 2003 Spatial correlations and higher-order gradient terms in a continuum description of dislocation dynamics *Acta Mater.* **51** 1271–81

[21] Wu R, Tüzes D, Ispánovity P D, Groma I, Hochrainer T and Zaiser M 2018 Instability of dislocation fluxes in a single slip: Deterministic and stochastic models of dislocation patterning *Phys. Rev. B* **98** 054110

[22] Stricker M, Sudmanns M, Schulz K, Hochrainer T and Weygand D 2018 Dislocation multiplication in stage II deformation of fcc multi-slip single crystals *J. Mech. Phys. Solids* **119** 319–33

[23] Groma I and Balogh P 1997 Link between the individual and continuum approaches of the description of the collective behavior of dislocations *Mater. Sci. Eng. A* **234–236** 249–52

[24] Rodney D, Le Bouar Y and Finel A 2003 Phase field methods and dislocations *Acta Mater.* **51** 17–30

[25] Kooiman M, Hütter M and Geers M G D 2014 Collective behaviour of dislocations in a finite medium *J. Stat. Mech. Theory Exp.* **2014** P04028

[26] El-Azab A 2000 Boundary value problem of dislocation dynamics *Model. Simul. Mater. Sci. Eng.* **8** 37–54

[27] El-Azab A 2000 Statistical mechanics treatment of the evolution of dislocation distributions in single crystals *Phys. Rev. B - Condens. Matter Mater. Phys.* **61** 11956–66

[28] Deng J and El-Azab A 2009 Mathematical and computational modelling of correlations in dislocation dynamics *Model. Simul. Mater. Sci. Eng.* **17** 075010

[29] Sedláek R, Schwarz C, Kratochvíl J and Werner E 2007 Continuum theory of evolving dislocation fields *Philos. Mag.* **87** 1225–60

[30] Arsenlis A, Parks D M, Becker R and Bulatov V V. 2004 On the evolution of crystallographic dislocation density in non-homogeneously deforming crystals *J. Mech. Phys. Solids* **52** 1213–46

[31]   Leung H S, Leung P S S, Cheng B and Ngan A H W 2015 A new dislocation-density-function dynamics scheme for computational crystal plasticity by explicit consideration of dislocation elastic interactions *Int. J. Plast.* **67** 1–25

[32]   Grilli N, Janssens K G F, Nellessen J, Sandlöbes S and Raabe D 2018 Multiple slip dislocation patterning in a dislocation-based crystal plasticity finite element method *Int. J. Plast.* **100** 104–21

[33]   Hochrainer T, Zaiser M and Gumbsch P 2007 A three-dimensional continuum theory of dislocation systems: Kinematics and mean-field formulation *Philos. Mag.* **87** 1261–82

[34]   Sandfeld S, Hochrainer T, Zaiser M and Gumbsch P 2011 Continuum modeling of dislocation plasticity: Theory, numerical implementation, and validation by discrete dislocation simulations *J. Mater. Res.* **26** 623–32

[35]   Sandfeld S and Zaiser M 2015 Pattern formation in a minimal model of continuum dislocation plasticity *Model. Simul. Mater. Sci. Eng.* **23** 065005

[36]   Monavari M and Zaiser M 2018 Annihilation and sources in continuum dislocation dynamics *Mater. Theory* **2** 3

[37]   Sudmanns M, Stricker M, Weygand D, Hochrainer T and Schulz K 2019 Dislocation multiplication by cross-slip and glissile reaction in a dislocation based continuum formulation of crystal plasticity *J. Mech. Phys. Solids* **132** 103695

[38]   Xia S, Belak J and El-Azab A 2016 The discrete-continuum connection in dislocation dynamics: I. Time coarse graining of cross slip *Model. Simul. Mater. Sci. Eng.* **24** 075007

[39]   Yefimov S, Groma I and Van der Giessen E 2004 A comparison of a statistical-mechanics based plasticity model with discrete dislocation plasticity calculations *J. Mech. Phys. Solids* **52** 279–300

[40]   Devincre B, Hoc T and Kubin L 2008 Dislocation Mean Free Paths and Strain Hardening of Crystals *Science (80-. ).* **320** 1745–8

[41]   Hochrainer T 2016 Thermodynamically consistent continuum dislocation dynamics *J. Mech. Phys. Solids* **88** 12–22

[42]   Dupuy L and Fivel M C 2002 A study of dislocation junctions in FCC metals by an orientation dependent line tension model *Acta Mater.* **50** 4873–85

[43]    Kurinnaya R, Zgolich M, Starenchenko V and Sadritdinova G 2016 The length change of a dislocation junction in FCC-single crystals under stress *AIP Conference Proceedings* vol 1698 (AIP Publishing LLC) p 040001

[44]    Madec R, Devincre B and Kubin L . 2002 On the nature of attractive dislocation crossed states *Comput. Mater. Sci.* **23** 219–24

[45]    Balluffi R W 2016 *Introduction to Elasticity Theory for Crystal Defects* (World Scientific Publishing Company)

[46]    Hirth J and Lothe J 1982 *Theory of Dislocations* (New York: John Wiley & Sons)

[47]    Hull D and Bacon D J 2011 *Introduction to dislocations* (Butterworth-Heinemann)

[48]    Kubin L P, Madec R and Devincre B 2003 Dislocation Intersections and Reactions in FCC and BCC Crystals *MRS Proc.* **779** W1.6

[49]    Higham D J 2008 Modeling and simulating chemical reactions *Psychother. Res.* **22** 1753–9

[50]    Deng J and El-Azab A 2010 Temporal statistics and coarse graining of dislocation ensembles *Philos. Mag.* **90** 3651–78

[51]    Zaiser M 2015 Local density approximation for the energy functional of three-dimensional dislocation systems *Phys. Rev. B* **92** 174120